\documentclass[prb,twocolumn,superscriptaddress,showpacs,aps,10pt,floatfix]{revtex4-1}

\usepackage{hyperref} 
\usepackage{graphicx} 
\usepackage{amsmath} 
\usepackage{amssymb}

\def\sgn{\mathop{\textrm{sgn}}} 
 
\newcommand{\beq}{
\begin{equation}
	} 
	\newcommand{\eeq}{
\end{equation}
} 
\newcommand{\be}{
\begin{eqnarray}
	} 
	\newcommand{\ee}{
\end{eqnarray}
} 
\newcommand{\Ref}[1]{Ref.~\onlinecite{#1}} 

\newcommand{\dg}{{\dagger}} 
 
\newcommand{\px}{{p_x}}

\newcommand{\bea}{
\begin{eqnarray}
	} 
	\newcommand{\eea}{
\end{eqnarray}
} 
 
\newcommand{\bK}{{\bf K}}

\begin{document}

\title{Graphene under spatially varying external potentials: Landau levels, magnetotransport, 
and topological modes} 
\author{Si Wu} \affiliation{Department of Physics, University of Toronto, Toronto, Ontario, M5S 1A7, Canada} \affiliation{Department of Physics and Astronomy, University of Waterloo, Waterloo, Ontario, N2L 3G1, Canada} 
\author{Matthew Killi} \affiliation{Department of Physics, University of Toronto, Toronto, Ontario, M5S 1A7, Canada} 
\author{Arun Paramekanti} \affiliation{Department of Physics, University of Toronto, Toronto, Ontario, M5S 1A7, Canada} \affiliation{Canadian Institute for Advanced Research, Toronto, Ontario, M5G 1Z8, Canada}
\affiliation{International Center for Theoretical Sciences, Bangalore 560 012, India}
\begin{abstract}
	Superlattices (SLs) in monolayer and bilayer graphene, formed by spatially periodic potential variations, lead to a modified bandstructure with extra finite-energy and zero-energy Dirac fermions with tunable anisotropic velocities. We theoretically show that transport in a weak perpendicular (orbital) magnetic field allows one to not only probe the number of emergent Dirac points but also yields further information about their dispersion. or monolayer graphene, we find that a moderate magnetic field can lead to a strong reversal of the transport anisotropy imposed by the SL potential, an effect which arises due to the SL induced dispersion of the zero energy Landau levels. This effect may find useful applications in switching or other devices. For bilayer graphene, we discuss the structure of Landau level wave functions and local density of states in the presence of a uniform bias, as well as in the presence of a kink in the bias which leads to topologically bound `edge states'. We consider implications of
these results for scanning tunneling spectroscopy measurements, valley filtering, and 
impurity induced breakdown of the quantum Hall effect in bilayer graphene. 
\end{abstract}


\maketitle
Graphene, a `Dirac material', exhibits such novel phenomena as an anomalous integer quantum Hall effect, weak anti-localization, and `zitterbewegung', which stem from its low energy excitations being massless chiral Dirac fermions. \cite{Novo2004,NetoRMP} In addition, it holds great potential for applications due to its high mobility and the experimental ability to tune its properties via gating. \cite{NetoRMP} Early experiments aiming to characterize the electronic properties of graphene used quantum Hall measurements to probe its underlying structure and unequivocally demonstrate the relativistic nature of the charge carriers \cite{zhang, novo2007}. In these studies, the Hall conductivity displayed plateaus at atypical values of $\pm 4 (e^2/h) (N+1/2)$, where $N$ is an integer. The origin of these plateaus is well known to be a consequence of the electrons in graphene being governed by a relativistic Dirac Hamiltonian, together with the inherent valley and spin degeneracy. The chiral symmetry of the Dirac Hamiltonian generates a particle-hole symmetric Landau level (LL) spectrum -- every positive energy LL has a conjugate negative energy LL that together are responsible for the positive and negative conductivity plateaus. The Dirac Hamiltonian also supports an additional zero energy LL, and leads to the `half-step' offset in the first conductivity plateaus. Finally, the step size between each plateau can be attributed to the presence of four degenerate Dirac cones labelled by different spin/valley indices. Further information about the velocity of the Dirac cone, $v_F$, can also be obtained by measuring the energy gap between the LL, which scales as $v_f \sqrt{B}$. Thus, information about LL spectrum and Hall conductivity of a system can provide direct evidence for the existence of Dirac-like quasiparticles and can be used to probe the degeneracy and velocity of the Dirac cones.

Inspired by studies of superlattices (SLs) in semiconductor systems \cite{Tsu}, such SLs have been proposed as a route to band structure engineering in graphene, paving the way to further new physics and applications. Recent studies of one dimensional (1D) superlattices (SLs) in monolayer graphene (MLG) have demonstrated their remarkable ability to not only regulate the Fermi velocity but also generate new Dirac cones. \cite{Park1,Park2,Park3,Park4,Brey,Barbier0,Barbier1,Barbier2,Barbier3,Killi1,Tan,Arovas,Burset,Plet,Rusponi} In seminal work, Park {\it et al.}, \cite{Park1} showed that a one dimensional (1D) SL in MLG could lead to a strong Fermi velocity renormalization in the direction transverse to the SL modulation, leading to strong transport anisotropies. Later, it was shown that SLs can even lead to nonzero energy Dirac points \cite{Park3} and multiple zero energy Dirac points near each valley. \cite{Park4,Brey} These effects stem from the chirality of the low energy quasiparticles in MLG. The new Dirac points generated by the SL lead to extra zero energy Landau levels for SLs in a weak orbital magnetic field. \cite{Park4}

For bilayer graphene (BLG), a minimal model of the low energy physics has quadratic band touching points which can be gapped out by an electric field perpendicular to the layers.\cite{McCann1,McCann2} 1D SLs in BLG are also predicted to display remarkable restructuring of the band dispersion, leading to new Dirac points in the spectrum which have anisotropic velocities determined by the SL potential, and are perturbatively stable against interactions (unlike the quadratic band touching point). For bilayer systems, there are two possible types of electrostatic SLs: (i) a chemical potential modulation where the potential profiles are the same on both layers, or (ii) an interlayer bias modulation where the layer potentials are of opposite sign leading to a periodically modulated electric field.

The effect of a chemical potential SL in BLG is highly dependent on the the strength of the modulation --- the band structure goes through a series of stages as the strength of the modulation is swept from weak to strong.\cite{Tan,Killi2} Starting from a weak potential, the quadratic band touching point splits into two anisotropic Dirac cones. This happens while preserving the net ``pseudospin winding number''.\cite{chpark2011} These Dirac points continually move out towards the mini-Brillouin zone (MBZ) before becoming gapped at a certain critical SL amplitude. \cite{Tan,Killi2} For even stronger SLs potentials, the widening of the band gap reverts and eventually closes by forming a series of new zero-energy Dirac points. The protection of the Dirac points has been shown to be directly related to the chirality of the low energy BLG quasiparticles.\cite{Tan,Killi2}

In the case of symmetric 1D electric field SLs in BLG, four anisotropic Dirac cones are generated, two at zero energy and two at finite energy. The existence of these Dirac points can be seen \cite{Killi2} to arise from a series of coupled `topological' modes \cite{Martin,Killi1,Killi2,Xavier} that emerge along the zero-line where the parity of interlayer bias reverses. Interestingly, these Dirac points are exceptionally robust to the size of the bias modulation. In the limit of a long period SL, these Dirac points evolve into independent topologically confined one-dimensional chiral modes.

In this paper, we investigate the electronic properties of both bilayer and monolayer graphene 1D SLs in the presence of a uniform magnetic field perpendicular to the plane.  For the bilayer, we consider two types of SLs: a chemical potential modulation and an electric field modulation. Similar to early magnetic field studies of graphene and graphene SLs \cite{zhang, novo2006, Park4}, careful analysis of the Landau levels spectrum and Hall conductivity proffers many potential experimental signatures of the underlying Dirac cones generated by the SL.  Special attention is also made to electric field modulations where the period length is large and decoupled `topological' modes emerge. Although our SL results are most relevant to graphene with chiral low energy excitations, the effects explored in this paper may also find interesting counterparts in low density two dimensional electron gas in semiconductor heterostructures, in 
which a honeycomb pattern has been successfully imprinted, with a possibility of other kinds of SL patterns imprinted in the future.\cite{Sinha}

We obtain the following key results for 1D SLs in MLG and BLG in a perpendicular (orbital) magnetic field. (1) We show that transport in a weak magnetic field can not only be used to probe for the presence of extra Dirac points, but may also be used, in clean systems, to obtain further information about their velocities via the energy and degeneracy of higher Landau levels (LLs). (2) We show that a moderate magnetic field in monolayer graphene leads to a dramatic reversal of the transport anisotropy generated by the SL potential alone, an effect arising from the SL induced dispersion of the zero energy LL. This field tunable transport anisotropy may find useful applications in monolayer graphene SLs. For example, one can use the change of the anisotropy as an on/off switch and even perform bit or gate operations with these SLs. However, this field tunable transport anisotropy is 
found to be absent in BLG. (3) We consider the Landau levels in BLG in the presence of a uniform bias, and a 
kink in the bias which leads to 1D topologically bound states, and the coupling between such  modes in an array of kinks. We also discuss the real space structure of the LL wave functions and 
the local density of states which is expected to be relevant to scanning tunneling spectroscopy (STS)
studies such as those described in Refs.~\onlinecite{connolly,rutter}. (4) Finally, we
consider possible implications of these results for valley filtering and breakdown of the quantum Hall effect.

For realistic SL period, $\lambda=300a$ with $a=1.42{\rm \AA}$ being the nearest neighbor lattice constant, the field strengths corresponding to weak and intermediate magnetic field are about 0.1T and 8T, respectively. Also, for this SL period, the minimum SL potential required to generate three Dirac points in one valley is about 200meV. All these values are experimentally achievable in order to see interesting physics described in points (1) and (2) above.

The organization of this paper is as follows:  Section \ref{section:mono} concerns  1D superlattices in monolayer graphene in a magnetic field.  We begin this section with a brief description of the band dispersion of 1D superlattices in the absence of a magnetic field.  In the next part, Section \ref{section:LL}, we introduce the low energy effective Hamiltonian and then identify the following three magnetic field regimes that lead to very different electronic dispersions: (i) weak, (ii) intermediate and (iii) high fields.  We present and the Landau level dispersions of these three regimes in Section~\ref{section:weak} - \ref{section:high}, respectively.  We then present the diagonal and Hall conductivity in Section~\ref{section:conductivity}. We find that the diagonal conductivity shows strong anisotropy reversal and Hall conductivity no longer exhibits plateaus, when magnetic field is tuned from weak to intermediate strength. 

Section \ref{section:bilayer} of this paper examines 1D superlattices in bilayer graphene and begins with description of the low energy theory that describes both chemical potential and electric field SLs.  We discuss the LL dispersion and transport for 1D chemical potential and electric field SLs in Sections~\ref{section:chemical} and \ref{section:electric}, respectively.

Then in the Section~\ref{section:realspace}, we consider the real space picture of interlayer modulations with dilute kinks and antikinks that form decoupled soliton modes.  We first provide an overview of the LL spectrum of uniformly biased BLG in Section~\ref{section:uniform} and then compare it to the spectrum generated when there is a single isolated antikink in the interlayer bias in the beginning of Section~\ref{section:kink}.  Also in this section, we present the tunneling current and provide key signatures that could identify the presence of the kink in STS measurements and suggest how the occurrence of disorder induced kinks may contribute to the breakdown of the quantum Hall effect.  Following the discussion of a single isolated interlayer bias kink, we discuss a periodic array of kink and antikinks in Section~\ref{section:array}.  We argue that for $\ell_B<\lambda$, it is only necessary to consider a single kink-antikink pair and observe that regardless of the magnetic field strength, the low energy physics is governed by the soliton modes which remain robust.  Employing a simple low energy effective theory to describe the soliton modes in a magnetic field, we gain further insights into the low energy physics of the SL and discuss why the zero energy LLs are robust from this perspective.  Finally, in Section~\ref{section:valley} we consider the coupling between the soliton states generated by a kink-antikink pair and how this leads to a single 1D one-way conducting band with definite valley index in each kink separately.  

\section{Monolayer graphene superlattices}\label{section:mono} 

\begin{figure}[t] 
	\includegraphics[width=.23\textwidth]{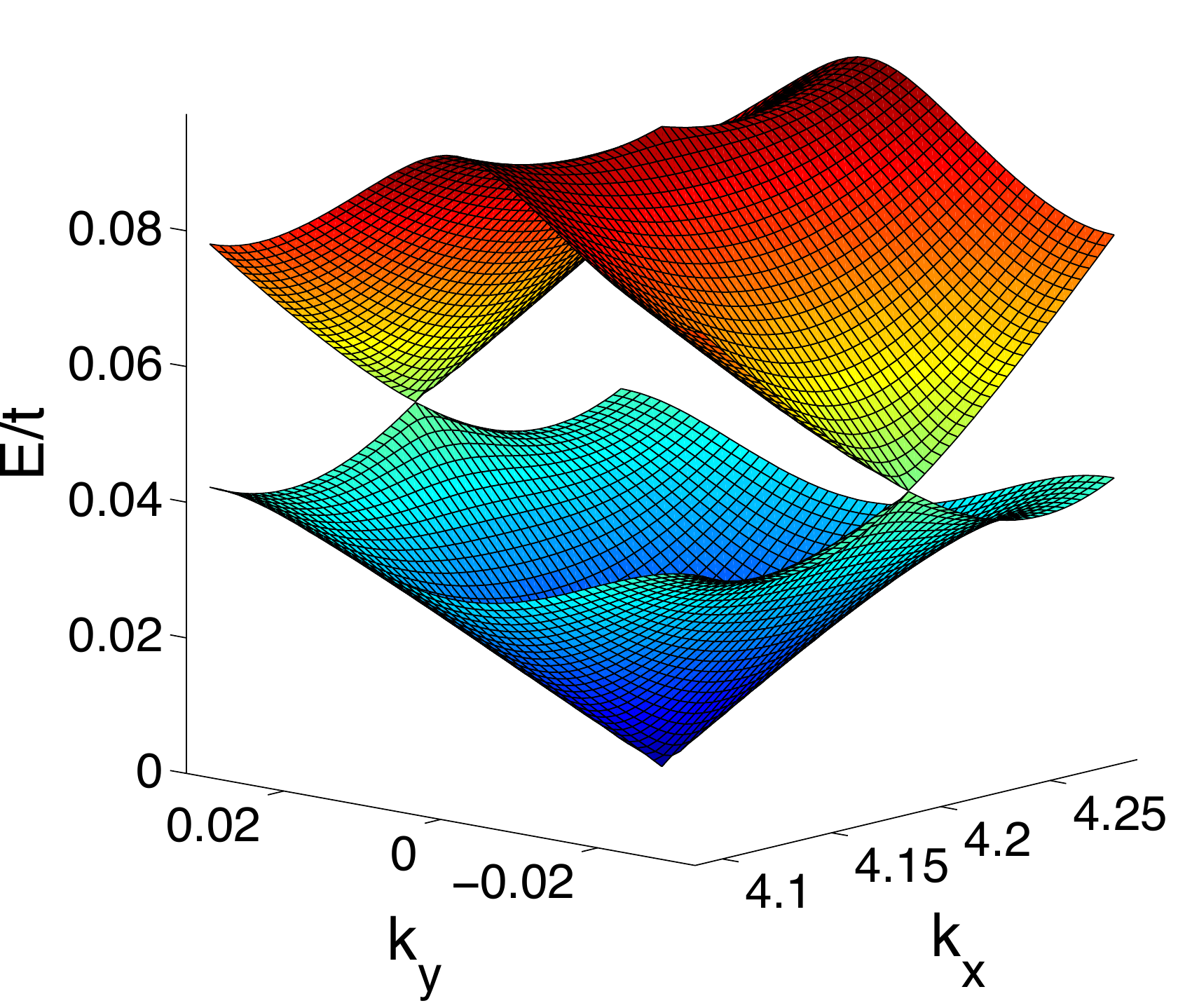} 
	\includegraphics[width=.23\textwidth]{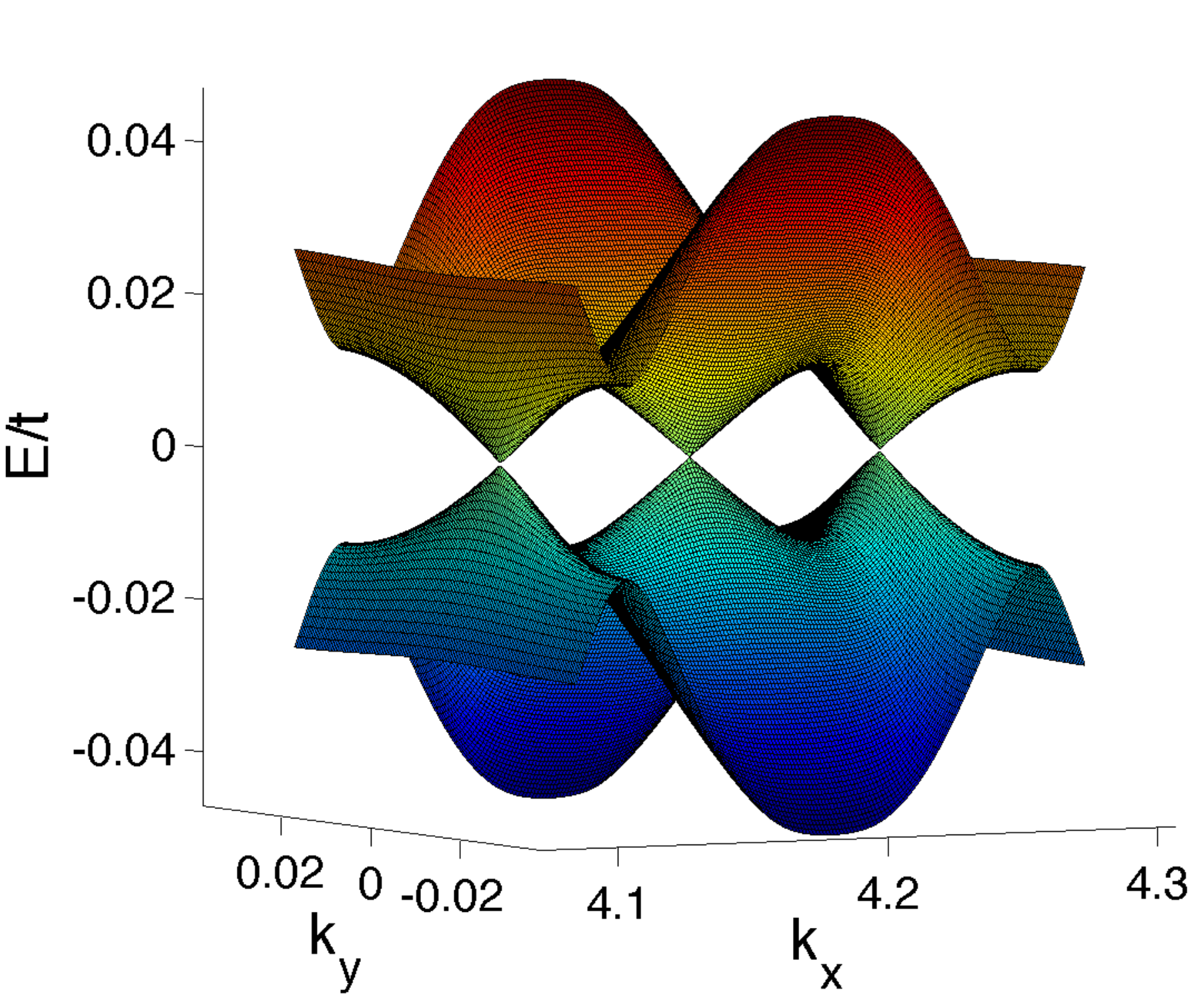}
	\caption{(Color online) Energy spectrum of monolayer graphene SLs with different (dimensionless) SL strengths, $\tilde{U}=1$ (left) and $\tilde{U}=3$ (right). For $\tilde{U}=1$, there is only one anisotropic zero energy Dirac cone present in the spectrum, while for $\tilde{U}=3$, two additional Dirac cones are generated in the direction perpendicular to SL. For even stronger SL potential, more Dirac cones will emerge in pairs.} 
	\label{GSL} 
\end{figure}

\subsection{Band structure of monolayer graphene superlattices}

It is useful to briefly review the effect of the SL potential \cite{Park4,Brey,Barbier2,Barbier3} at zero magnetic field on the low energy spectrum near $\bK$ before we present the SL spectrum in a nonzero magnetic field (see Fig. \ref{GSL}). For $U=0$, where $U$ is the SL strength, the spectrum consists of a single Dirac point at $\bK$, with an isotropic velocity at low energy. The first effect of turning on a nonzero $U$ is to make this Dirac cone anisotropic, with $v_y = v_f$, but $v_x < v_f$. For $U = 4\pi v_f/\lambda$, with $\lambda$ as the SL period, one finds $v_x = 0$. Further increasing $U$, with $4\pi v_f/\lambda < U < 8\pi v_f/\lambda$, leads to {\it three} zero energy Dirac points: one of these continues to be located at $\bK$, while two new Dirac points emerge which are symmetrically split off from the $\bK$ point along the $\pm x$-direction. All three Dirac cones have anisotropic velocities.\cite{Barbier2} It is clearly convenient to measure the strength of the SL potential in terms of a dimensionless parameter $\tilde{U}$, defined via $U \equiv (2 \pi v_f/\lambda) \tilde{U} $, so that tuning $\tilde{U}$ with $0 < \tilde{U} < 4$ leads to the different interesting spectra discussed above. A further increase in $\tilde{U}$ leads to a more complex spectrum with even more Dirac points. 

\subsection{Landau Level spectrum}\label{section:LL}

Ignoring spin, the low energy Hamiltonian of pristine MLG is given by a $2\times 2$ matrix at each valley, $ \hat{H}_0 \! = \! v_f (s \pi_x \sigma_x - \pi_y \sigma_y)$, where pseudospin $\sigma_z=\pm 1$ labels the two sublattices, while the two (decoupled) valleys at $\pm \bK=\pm 4\pi \hat{x}/3$ are labelled by $s \! = \! \pm 1$. Here, $v_f \! = \! 3ta/2$ is the isotropic Fermi velocity, with $a \! = \! 1.42\text{\text{\AA}}$ and $t \! = \! 3$eV being the nearest neighbor carbon-carbon distance and transfer integral respectively. (We set $\hbar \! =\! 1$ for convenience.) In a uniform perpendicular magnetic field, $\pi_j \! = \! -i \nabla_j \! - \! e A_j$; for ${\bf B} \! =\! - B \hat{z}$, and in the Landau gauge, the vector potential ${\bf A} \! = \! By\hat{x}$.

Diagonalizing $\hat{H}_0$, we find energies $\varepsilon_n={\rm sgn}(n)\sqrt{|n|}\omega_c$, where $\omega_c \! =\! \sqrt{2}v_F/\ell_B$, with $\ell_B \! =\! 1/\sqrt{eB}$. For $s=+1$ (i.e., at valley ${\bK}$), the $n \! \neq \! 0$ eigenfunctions are given by \be &&\phi_{n,k,+} (x,y)=\frac{e^{ i k x}}{\sqrt{2 L}}\left( 
\begin{array}{c}
	\psi_{|n|,k}(y) \\
	- {\rm sgn}(n) \psi_{|n|-1,k}(y) 
\end{array}
\right), \label{eigen1} \ee where $L$ and $k$ are the system length and electron momentum deviation from $\bK$, both along the $x$-direction, while for $n=0$, \beq \phi_{0,k,+} (x,y)=\frac{e^{ikx}}{\sqrt{L}} \left( 
\begin{array}{c}
	\psi_{0,k}(y) \\
	0 
\end{array}
\right). \label{eigen2} \eeq Here, $\psi_{n,k}(y)$ is the n-th eigenstate of a (shifted) 1D harmonic oscillator, \be \psi_{n,k}(y)&=&\frac{1}{\sqrt{2^{n}n!\sqrt{\pi}\ell_B}}{\rm exp} \left[-\frac{1}{2}\left(\frac{y-y_0}{\ell_B}\right)^2\right]\nonumber\\
&&\times H_{n}\left(\frac{y-y_0}{\ell_B}\right), \ee centered at $y_0 \! = \! k\ell_B^2$, and $H_n$ are Hermite polynomials. For $s\!=\!-1$ (i.e., at $- \bK$), the eigenfunctions are given by $\phi_{n,k,-}(x,y) \!=\! -i \sigma_y \phi_{n,k,+}(x,y)$. The full low energy LLs of MLG are thus $\phi_{n,k,\pm} (x,y) {\rm e}^{\pm i K_x x}$.

We now turn to the effect of a periodic 1D chemical potential modulation $V(y)$, with period $\lambda \! \gg \! a$, on these Landau levels. The set of eigenfunctions $\phi_{n,k,s} (x,y) {\rm e}^{ i s K_x x}$, with $s=\pm 1$, form a convenient basis to study the SL Hamiltonian in a magnetic field. (This basis choice is different from the one used by Park, {\it et al}, and allows us to numerically access a wide range of magnetic fields.) Due to momentum conservation along the $x$-direction, the SL Hamiltonian is diagonal in $k$. Further, for $\lambda \gg a$, intervalley scattering is strongly suppressed. We will therefore assume that the two valleys stay completely decoupled. (We focus below on valley ${\bK}$ with $s\!=\! +1$; we expect identical physics around valley $-{\bK}$.) With this approximation, the only effect of the SL potential is, thus, to induce Landau level mixing.

To proceed, we need to choose a concrete form for the SL potential. For simplicity, we set $V(y) \! =\! \frac{U}{2}\cos\left(\frac{2\pi y}{\lambda}\right)$, although our results can be easily generalized to other SL potentials which are smooth on the atomic scale by including multiple Fourier components. For numerical computations, the following integral identity \cite{DLMF} for harmonic oscillator states proves useful: 
\be
&& \int_{-\infty}^{\infty} dy \cos\left(\frac{2\pi y}{\lambda}\right)\psi_{n,k}(y) \psi_{n+m,k}(y) \nonumber \\
= && \left(\frac{n!}{(n+m)!}\right)^{1/2} \!\! \left(\frac{\sqrt{2}\pi\ell_B}{\lambda}\right)^m \!\! \exp\left(\!-\frac{\pi^2\ell_B^2}{\lambda^2}\! \right) \nonumber\\
&&\times  L_n^{(m)}\left(\frac{2\pi ^2\ell_B^2}{\lambda ^2}\right) \cos\left(\frac{2\pi y_0}{\lambda}+\frac{m\pi}{2}\right), \label{identity} \ee where $m \geq 0$, and $L_n^{(m)}$ is the generalized Laguerre polynomial. Using this identity, we simplify and numerically compute the matrix elements of the full Hamiltonian, retaining up to 3000 Landau levels, and diagonalize this to obtain the spectrum of the 1D SL in a magnetic field. 

In order to study the effect of the magnetic field on the 1D SL in graphene, with $\tilde{U} \sim {\cal O}(1)$, it is useful to consider three regimes for the magnetic field.
\begin{figure}[t] 
	\includegraphics[width=.48
	\textwidth]{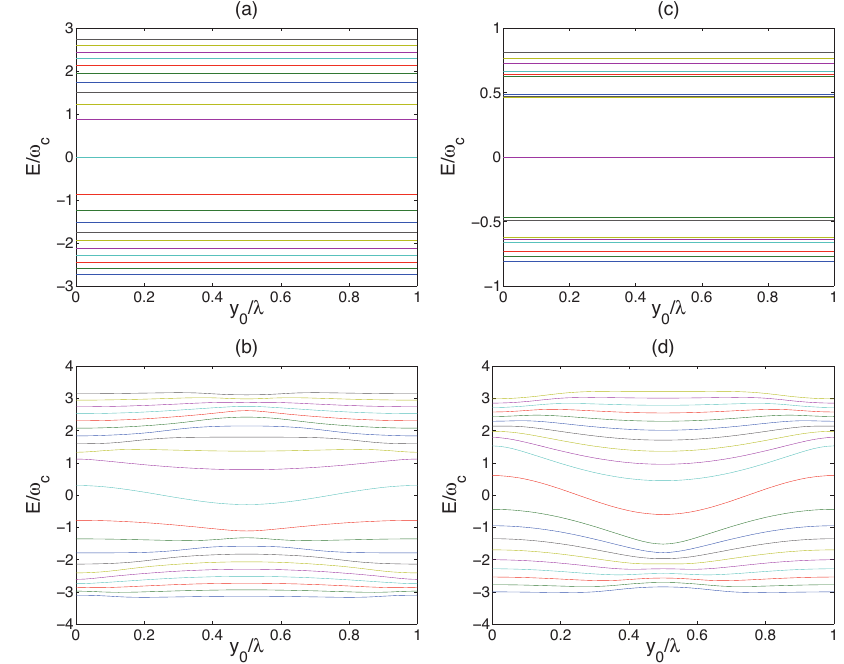} 
	\caption{(Color online) Landau levels of monolayer graphene SL for different (dimensionless) SL strengths $\tilde{U}$, and magnetic fields $B$. The spectrum is shown for weak field ($\ell_B=2\lambda$, top panels) and intermediate field ($\ell_B=0.2\lambda$, bottom panels). Left panels (a,b) correspond to $\tilde{U}=1$ which supports a single anisotropic zero energy massless Dirac fermion. Right panels (c,d) correspond to $\tilde{U}=3$ which supports three zero energy massless Dirac fermions with anisotropic velocities - the weak field zero energy LL thus has three times as many states for $\tilde{U}=3$ as it does for $\tilde{U}=1$, while the $n=\pm 1, \pm 2$ levels have degeneracy splitting in weak field due to the Dirac fermions having two different mean velocities. For $\ell_B \ll \lambda$ (not shown), the LLs closely resemble that of pristine graphene. See text for a detailed discussion of the Landau level structure.} 
	\label{GSL_LL} 
\end{figure}

(i) {\bf Weak field}: This regime corresponds to having $\hbar \omega_c \ll U$, where the Landau level spacing is much smaller than the SL amplitude, so that $2 \pi \ell_B/\lambda \gg 1$. In this regime, the magnetic field may be viewed as effectively `probing' the zero field SL excitations.

(ii) {\bf Intermediate field}: In this regime, $\hbar \omega_c \sim U$, which means $2\pi \ell_B/\lambda \sim 1$, so that the SL potential and the magnetic field have to be treated on equal footing.

(iii) {\bf Strong field}: Here, $\hbar \omega_c \gg U$ or, equivalently, $2\pi \ell_B/\lambda \ll 1$. In this regime, the SL potential only weakly perturbs the Landau levels of pristine graphene.

Fig.~\ref{GSL_LL} shows the spectrum of the graphene SL in different field regimes for SL strengths $U\!\!=\!\!2\pi v_f/\lambda$ (or $\tilde{U}=1$) and $6\pi v_f/\lambda$ (or $\tilde{U}=3$). This allows us to contrast the behaviour of the spectrum of the SL in a magnetic field without or with extra Dirac points being present at zero field, and to explore consequences for quantum Hall physics and transport.

\subsubsection{Weak field regime} \label{section:weak}

When the magnetic field is weak, $\ell_B=2\lambda$ (top panels in Fig. \ref{GSL_LL}), we find that the energy spectrum barely depends on the value of $k$, or equivalently, $y_0$. This is due to the fact that when magnetic length $\ell_B$ is larger than the SL period $\lambda$, the matrix elements of the Hamiltonian do not depend on the center of the LL wavefunctions, which yields flat bands. Equivalently, in this regime, the magnetic field may be viewed as effectively `probing' the structure of the zero field SL dispersion leading to Landau levels which depend on the nature of the Dirac spectrum at low energy.

For $\tilde{U}=1$, the low energy spectrum of the SL contains a single anisotropic Dirac point at zero energy. For an anisotropic Dirac cone described by an effective Hamiltonian $H=v_xk_x\sigma_x+v_yk_y\sigma_y$, the LLs are given by $\varepsilon_n=\sgn(n)\sqrt{2|n|v_xv_y}/\ell_B$. Since the SL renormalizes $v_x < v_f$, but leaves $v_y=v_f$, the Landau levels at weak field resemble those of pristine graphene, but with a renormalized lower effective velocity $\sqrt{v_x v_y} < v_f$.

For $\tilde{U}=3$, the low energy spectrum of the SL contains three anisotropic Dirac points at zero energy, so that the zero energy Landau level has three times the degeneracy of the case with $\tilde{U}=1$. Further, the Dirac cone centred at $\bK$ has a slightly different average velocity $\sqrt{v_x v_y}$ compared with the two cones which are symmetrically split off from $\bK$ along $\pm \hat{x}$. This degeneracy breaking results in the Landau levels at nonzero energy becoming weakly split, as is most clearly seen for the first two excited Landau levels (at positive or negative energy, i.e., with $n=\pm 1, \pm 2$). We have numerically determined $v_x$ and $v_y$ for each of the three Dirac points and found good agreement between the energy levels obtained on this basis of having Dirac fermions with two different average velocities, and that obtained directly numerically.

At higher energies, $E/\omega_c \gtrsim 2$ for $\tilde{U}=1$ or $E/\omega_c \gtrsim 1$ for $\tilde{U}=3$, the spectrum begins to deviate from this simple behavior expected for a linear Dirac spectrum. This deviation results from curvature in the dispersion, which appears upon going beyond the linearized approximation.

\subsubsection{Intermediate field regime} 
\label{section:inter}

At intermediate fields, for $\ell_B=0.2\lambda$, the spectrum at low energy is most simply understood as arising from the SL potential inducing a strong dispersion to the Landau levels. In simple terms, if we assume that the state labelled by momentum $k$, or equivalently position $y_0$, have an energy which is modulated by the SL potential, we expect a periodic modulation of this energy with period $\lambda$ and amplitude proportional to the SL amplitude $U$. The behaviour of the low energy Landau levels, $n=0,\pm 1, \pm 2$, as seen from the lower panels in Fig.\ref{GSL_LL}, is consistent with this scenario, with the modulation following the $\cos(2\pi y/\lambda)$ form of the SL potential and the modulation for $\tilde{U}=3$ being roughly thrice as strong as the modulation for $\tilde{U}=1$. We can also see that the low energy Landau levels when $\tilde{U}=3$ overlap with each other. This will have nontrivial effect on the dc conductivity, as shown in the following subsection. For higher energy Landau levels, the energy spectrum still has a periodic modulation but no longer retains the simple form of cosine function. This is due to the fact that as the energy gets higher, the distribution of Landau levels becomes more dense and the energy difference between two adjacent levels is now comparable to the matrix element of SL potential. Therefore, a simple first order perturbation correction is not enough to account for the dispersion and second order perturbation from adjacent levels must be taken in account, which causes the Landau level loses the simple cosine form.

\subsubsection{High field regime}
\label{section:high}

For very strong magnetic field, the Landau level structure of pristine graphene is recovered. Here, only one zero energy level exists at the Dirac point, and other energy levels follow the square root relation. This is simply because in such a strong magnetic field, the SL is just a perturbation and can only give rise to a small modulation of the LLs following our argument at intermediate field. From a perturbative point of view, the energy corrections up to first order to the LL energies are given by \beq \Delta E^{(1)}=\int dy \phi_n^*(k,y)V(y)\phi_n(k,y), \eeq which, upon using the identity (\ref{identity}), gives a sinusoidal dependence on the center position of LL wavefunctions. Thus, even in a strong magnetic field, the energy spectrum is not dispersionless but has a spatial modulation following the SL; however the ratio of the amplitude of this modulation to the Landau level spacing, $U/\omega_c$, is extremely small in the high field regime. This dispersion, though small, can give rise to interesting magnetoresistance oscillation known as Weiss oscillation, on top of the usual Shubnikov-de Hass oscillation. \cite{Matulis} It was shown that, compared to two-dimensional electron gas with parabolic dispersion relation, Weiss oscillation in graphene SL is more pronounced and is more robust against temperature damping in small field region. This is a consequence of the different Fermi velocities of Dirac and normal electrons at same chemical potential. \cite{Matulis} Similarly, Weiss oscillation in bilayer graphene chemical potential SLs has also been discussed. \cite{Tahir}
\begin{figure}
	[t] \centering 
	\includegraphics[width=.48
	\textwidth]{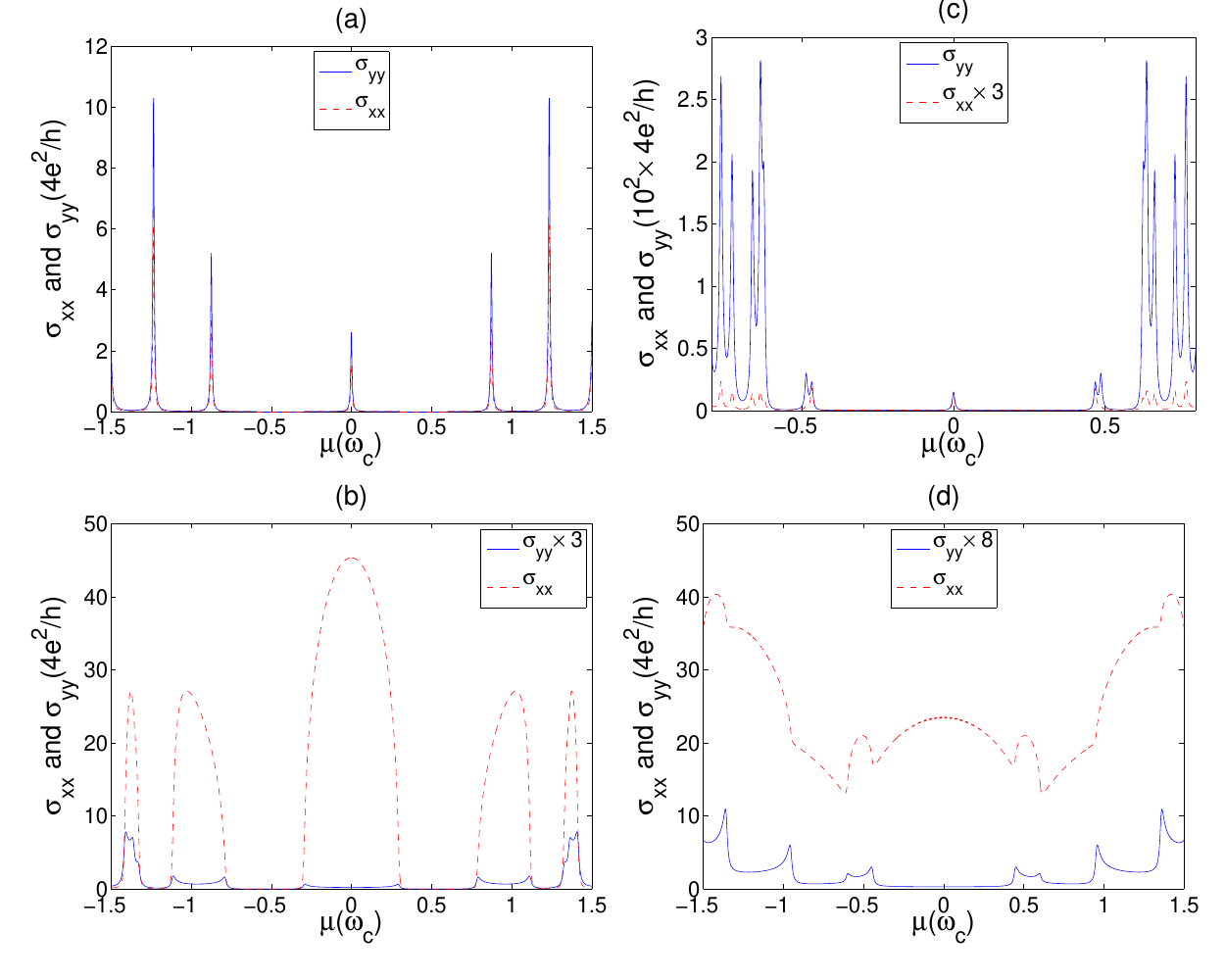}  \caption{ (Color online) Diagonal dc conductivities of monolayer graphene SL for different (dimensionless) SL strengths $\tilde{U}$, and magnetic fields $B$. The conductivity is shown for weak field ($\ell_B=2\lambda$, top panels) and intermediate field ($\ell_B=0.2\lambda$, bottom panels). Left panels (a,b) correspond to $\tilde{U}=1$, and right panels (c,d) correspond to $\tilde{U}=3$. The conductivities show strong anisotropy when magnetic field strength is tuned - for weak field (a,c), $\sigma_{yy}$ is larger than $\sigma_{xx}$, which is a consequence of the Fermi velocity renormalization in the absence of magnetic field; for moderate field (b,d), the anisotropy is reversed, since $\hat{v}_x$ acquires intra-LL contributions, as explained in the text. For $\ell_B \ll \lambda$ (not shown), result for pristine graphene is recovered and the transport becomes isotropic.} \label{diagonal_SLG} 
\end{figure}

\subsection{Diagonal and Hall Conductivity in MLG SLs}
\label{section:conductivity}

Once we have the eigenvalues and eigenfunctions for the superlattice in a perpendicular magnetic field, both ac and dc conductivities can be calculated directly by Kubo formula, \cite{Meso} 
\be 
\sigma_{ij}(\omega)&=&\frac{iv_F^2}{\lambda\ell_B^2}\int_0^{\lambda}dy_0\sum_{\alpha,\beta} 
\frac{f(E_{\alpha})-f(E_{\beta})}{E_{\alpha}-E_{\beta}}\nonumber\\
&&\times \frac{\langle\alpha k |v_i|\beta k\rangle\langle\beta k |v_j|\alpha k\rangle}
{E_{\alpha}-E_{\beta}-\omega-i\gamma}. 
\label{Kubo}
\ee  Here, we have set $\gamma=10^{-3}\times \omega_{c}$ as the Landau level broadening and measure the conductivities in the unit of $e^2/h$, $E_{\alpha}(y_0)$ and $|\alpha k\rangle$ are the $\alpha$-th eigenvalue and the corresponding eigenstate of the system which can be expanded in the basis of $|nk\rangle$, with $y_0=k\ell_B^2$ and $\phi_n(k,y)=\langle y|nk\rangle$ being the LL wavefunctions for pristine graphene. $f(E)$ is the Fermi-Dirac distribution, and $v_i$ is the velocity operator in $\hat{i}$-direction and $v_i=v_F\sigma_i$, where $\sigma_i$ is the Pauli matrix. Using Eq. (\ref{eigen1}), (\ref{eigen2}), we can get the expression for the matrix elements of velocity operators from \be \langle m k |v_x| n k\rangle &=& v_F\big[{\rm sgn}(m)\delta_{|m|-1,|n|}\nonumber\\
&& +{\rm sgn}(n)\delta_{|m|,|n|-1}\big], \ee and \be \langle m k |v_y| n k\rangle &=& -iv_F\big[{\rm sgn}(m)\delta_{|m|-1,|n|}\nonumber\\
&& -{\rm sgn}(n)\delta_{|m|,|n|-1}\big], \ee by expanding $|\alpha k\rangle$ into $|nk\rangle$. Note that $\langle \alpha k|v_y|\alpha k\rangle=0$ is always true for any state. This is because in the $\hat{y}$ direction, the wavefunction is always localized.

Results for dc diagonal conductivities as a function of chemical potential $\mu$ are shown in Fig. \ref{diagonal_SLG}. This can be done by setting the frequency $\omega$ to zero in Eq. (\ref{Kubo}), and only the real part of the conductivity tensor is nonzero. In weak magnetic field, the conductivities show strong anisotropy, with $\sigma_{yy}$ larger than $\sigma_{xx}$, which is a direct consequence of the Fermi velocity renormalization in the absence of magnetic field (see Fig. \ref{diagonal_SLG} (a) and (c)). Besides, since $\langle \alpha k|v_y|\alpha k\rangle=0$ and $\langle \alpha k|v_x|\alpha k\rangle\simeq 0$ due to the flat band structure, the major contribution to the diagonal conductivities comes from off-diagonal matrix elements, $\langle \alpha k|v_i|\beta k\rangle$ with $\alpha\neq\beta$. Since the matrix elements of $v_y$ are always larger than those of $v_x$ due to the anisotropic dispersion, this gives rise to the anisotropy in the weak field. In intermediate magnetic field, conductivities still show anisotropy, but with $\sigma_{xx}$ significantly larger than $\sigma_{yy}$ (see Fig. \ref{diagonal_SLG} (b) and (d)). This is because $v_x$ has acquired diagonal matrix element, $\langle \alpha k|v_x|\alpha k\rangle=
\partial E_{\alpha}(y_0=k\ell_B^2)/
\partial k\neq 0$ since the energy spectrum is dispersive, while $v_y$ still lacks this contribution. Notice the positions of the conductivity peaks of $\sigma_{yy}$ exactly correspond to the minimum and maximum of the energy band, where the density of states is the largest. For $\sigma_{xx}$, however, the conductivity is minimum at the band edge, since the average of the velocity operator, $\langle v_x \rangle$, is zero at these energies. Therefore, the intra-LL contribution to $\sigma_{xx}$ is the smallest at the band edge and thus leads to dips in it. For weak SL potential, $\sigma_{xx}$ can drop to zero when there is no overlapping LLs, while in a strong SL, $\sigma_{xx}$ always show dispersive transport property. In strong magnetic field (not shown here), where the Landau levels structure of pristine graphene is recovered, the conductivities become isotropic (see, for example, Ref. \onlinecite{Ferreira}). In this case, the SL is merely a perturbation to the magnetic field and thus should have minor effect on determining the magnetotransport properties.

\begin{figure}
	[t] \centering 
	\includegraphics[width=.48
	\textwidth]{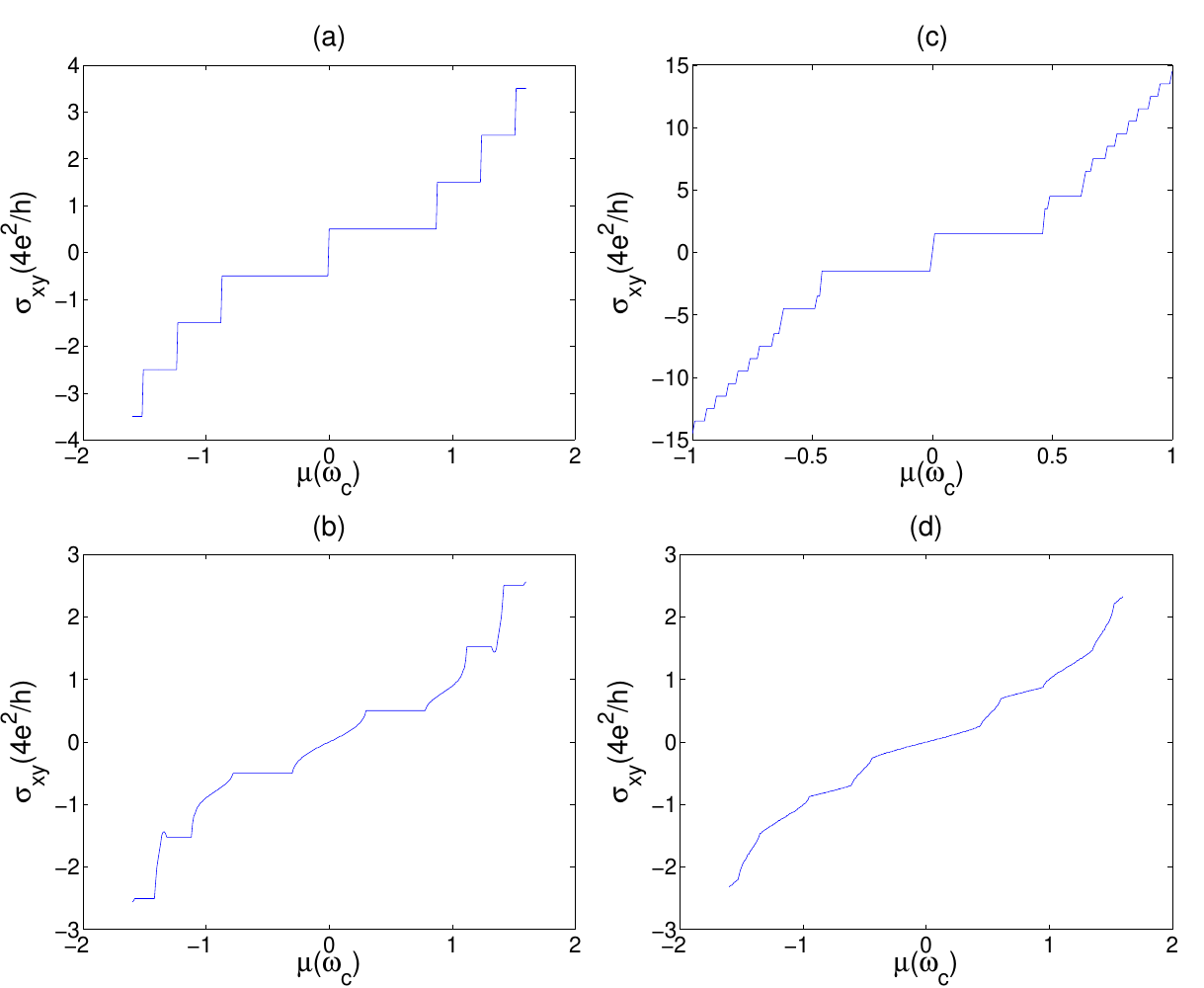}  \caption{ (Color online) The dc Hall conductivity of monolayer graphene SL for different (dimensionless) SL strengths $\tilde{U}$, and magnetic fields $B$. The conductivity is shown for weak field ($\ell_B=2\lambda$, top panels) and intermediate field ($\ell_B=0.2\lambda$, bottom panels). Left panels (a,b) correspond to $\tilde{U}=1$, and right panels (c,d) correspond to $\tilde{U}=3$. For weak field (a,c), the Hall conductivity shows well-defined plateaus, as a consequence of nearly flat energy bands. For intermediate field (b,d), the energy bands become dispersive and the Hall conductivity no longer shows step-like structure. However, for weak SL (b), the energy bands are not fully overlapped, Hall conductivity still shows small plateaus when chemical potential falls between two bands, and the value of $\sigma_{xy}$ changes by one between adjacent steps, as expected from Dirac physics. For $\ell_B \ll \lambda$ (not shown), result for pristine graphene is recovered and Hall conductivity is constant between adjacent LLs and changes by one when chemical potential crosses a LL.} \label{Hall_SLG} 
\end{figure}

The dc Hall conductivity is shown in Fig. \ref{Hall_SLG}. For weak magnetic field (Fig. \ref{Hall_SLG} (a) and (c)), the Hall conductivity shows well-defined plateaus, as a consequence of nearly flat energy bands. The values of Hall conductivity around Dirac points, taking into account of spin and valley degeneracies, are $\pm 1/2(4e^2/h)$ in weak SL ($\tilde{U}=1$) and $\pm 3/2(4e^2/h)$ in strong SL ($\tilde{U}=3$).  This result resembles the anomalous half integer quantum Hall effect in pristine graphene and the Hall conductivity triples due to the existence of three Dirac points in a strong SL. Moving away from the Dirac point, we can observe quantum Hall plateaus with higher conductivities, and the value increases by 1 each time the chemical potential crosses an LL. For intermediate magnetic field (Fig. \ref{Hall_SLG} (b) and (d)), there is no longer well defined plateaus due to the dispersive energy spectrum. However, for weak SL, the LLs are not overlapped with each other. If chemical potential falls between two LLs, a small plateau can still show up, with the value expected from Dirac physics. When the magnetic field becomes strong enough as the LL structure for pristine graphene is restored, Hall conductivity will show anomalous half integer quantum Hall plateaus.

\begin{figure}[t] 
	\centering 
	\includegraphics[width=.48\textwidth]{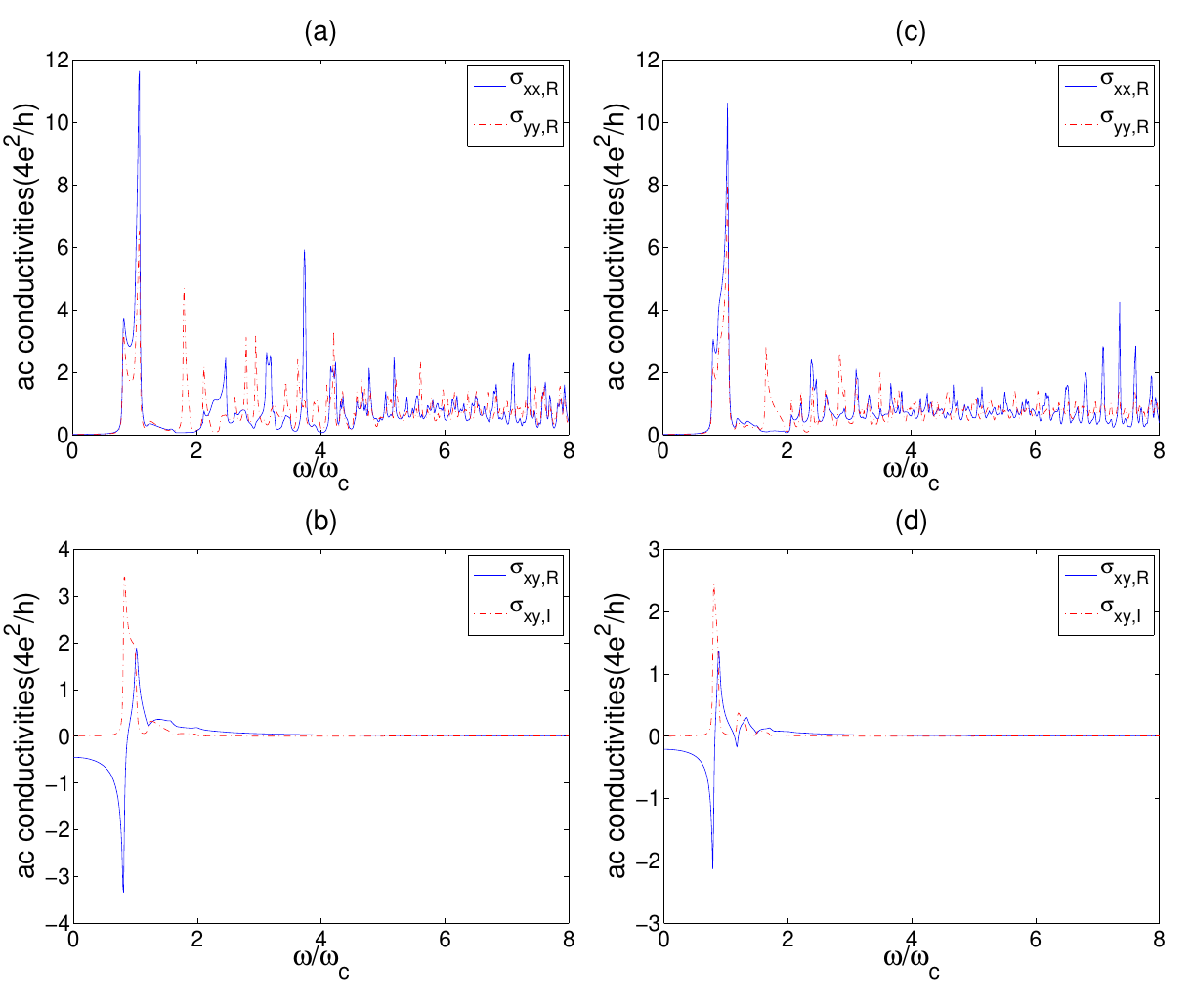} 
	\caption{ (Color online) The ac conductivity of monolayer graphene SL for different (dimensionless) SL strengths $\tilde{U}$, in intermediate magnetic field, $\ell_B=0.2\lambda$, with $\mu=0.2\omega_c$. Left panels (a,b) correspond to $\tilde{U}=1$, and right panels (c,d) correspond to $\tilde{U}=3$.} 
	\label{ac_SLG} 
\end{figure}

Fig. \ref{ac_SLG} shows the ac conductivities of graphene SLs in an intermediate magnetic field. For weak and strong 
magnetic fields, the results resemble those of pristine graphene,\cite{Ferreira} since in both cases the LLs are nearly 
flat and the real part of the conductivities show strong peaks when photon frequencies exactly correspond to the energy 
differences between two LLs. In an intermediate magnetic field, the result is complicated by the dispersion of LLs. At 
low frequencies, there can be optical transitions in a range of photon energies, and the real part of diagonal 
conductivities is maximum at the band edge where the DOS is also maximum. At high frequencies, the LLs become less 
dispersive and peaks will show up. These results can be linked with graphene's unusual magneto-optical properties, 
for example, giant Faraday rotation.\cite{Ferreira, Crassee} 
While the anisotropy in the diagonal conductivities can lead to 
anisotropic rotation angles for incident waves with different polarization plane, this effect is actually quite small 
and hard to observe experimentally.

\section{Bilayer Graphene Superlattices}\label{section:bilayer}

\begin{figure}[t] 
	\includegraphics[width=.23\textwidth]{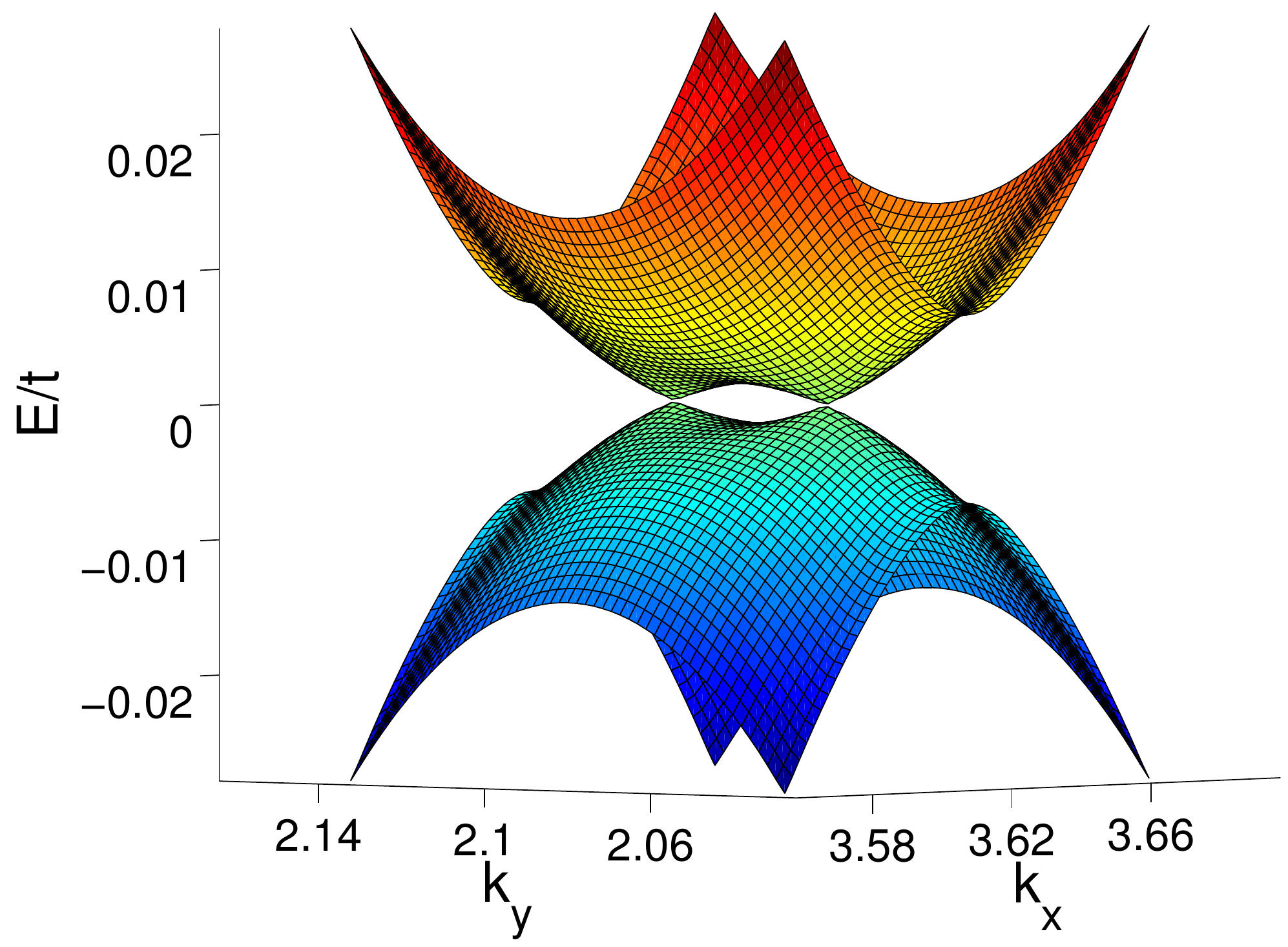} 
	\includegraphics[width=.23\textwidth]{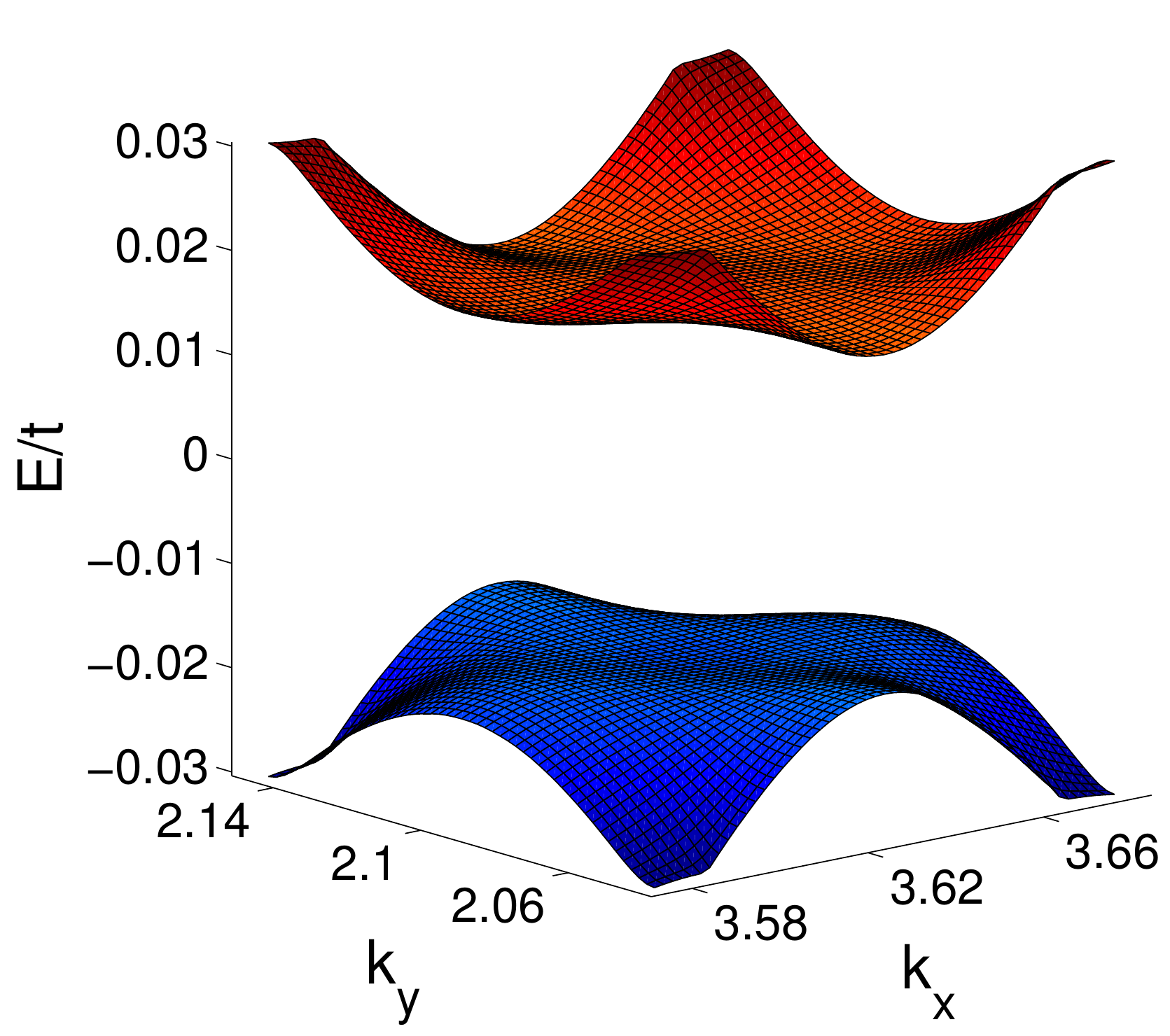}
	\includegraphics[width=.23\textwidth]{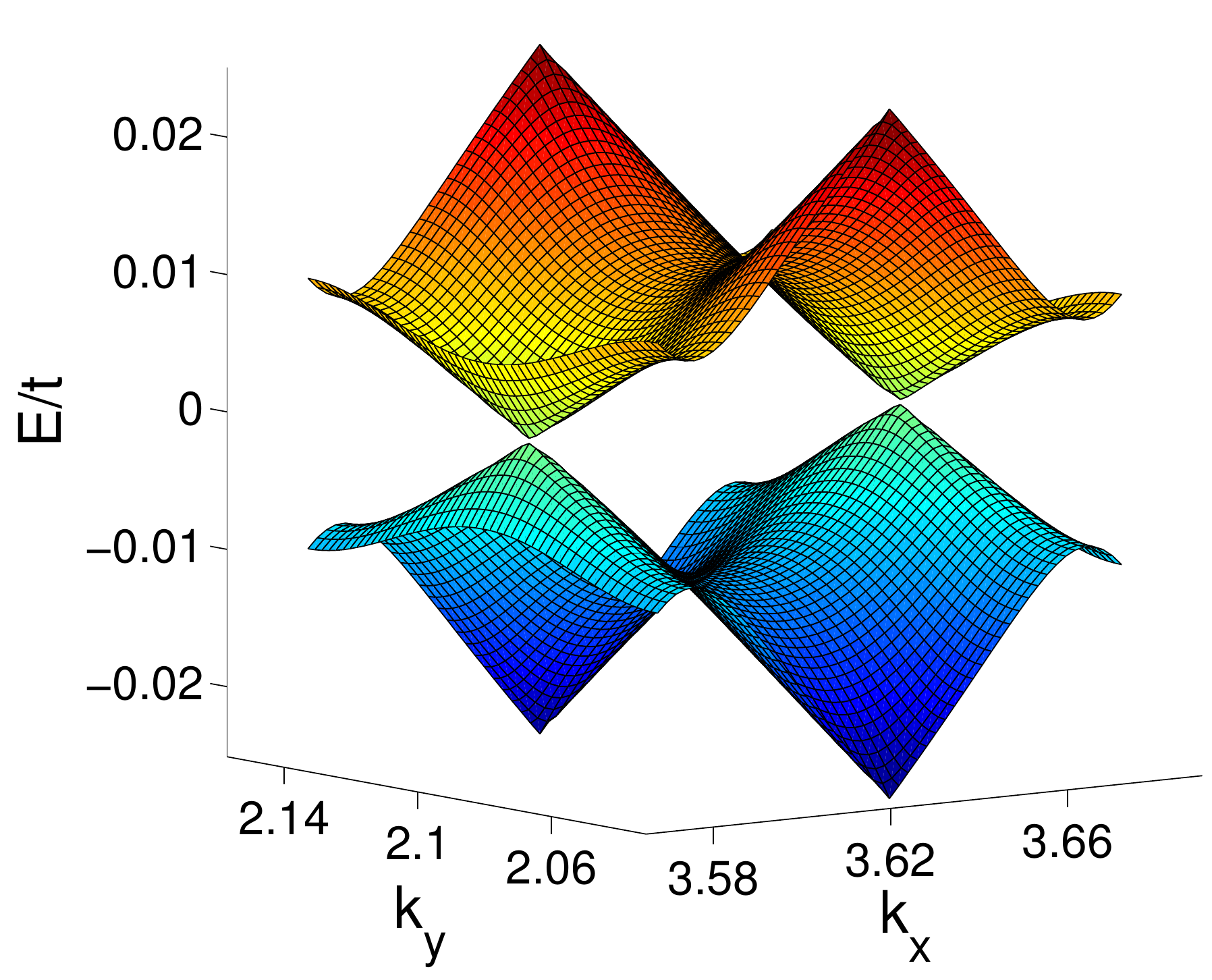} 
	\includegraphics[width=.23\textwidth]{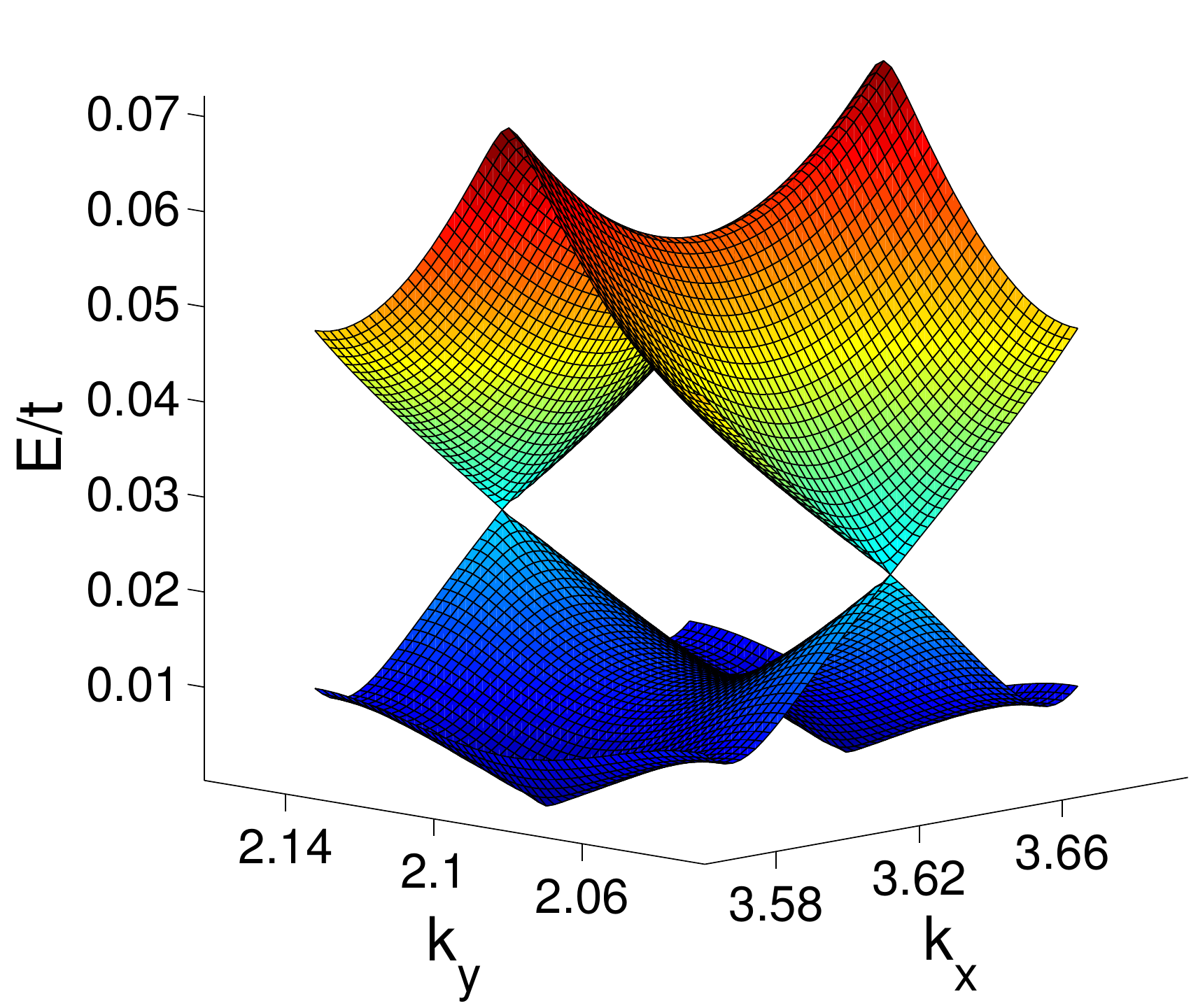}
	\caption{(Color online) Band structures for different types of BLG SLs. In a chemical potential SL, when the SL strength is weak ($V_0=0.01t$, top left), two anisotropic Dirac cones are generated in the spectrum. When the SL strength exceeds a critical value, the spectrum becomes completely gapped out ($V_0=0.04t$, top right). In an electric field SL, however, there are always two Dirac cones at zero energy ($V_0=0.04t$, bottom left). In addition to the zero energy Dirac cones, there are finite energy Dirac cones (bottom right). These results can be understood from chirality protected level crossing (for chemical potential SL) and an effective ``wire'' model (for electric field SL) presented in \Ref{Killi1}.} 
	\label{BLGSL} 
\end{figure}

\subsection{Band structure of bilayer graphene superlattices}

Before presenting the results for BLG SLs in a perpendicular magnetic field, we briefly review the band structures of BLG SLs in the absence of a magnetic field. In BLG, there are two distinct types of SLs, namely, chemical potential and electric field SLs. Other SLs can be thought as a 
superposition of these two fundamental ones, and can be similarly studied.

In a chemical potential SL, the SL potentials are exactly the same on each layer. As has been shown by us in previous work, \cite{Killi1} the electron and hole state are completely decoupled along the SL direction. Therefore, the SL correction to the energy spectrum pushes electron (hole) states down (up) and causes the two quadratic bands cross each other inside the mini Brillouin zone (MBZ). However, in other directions, this decoupling is absent and level anti-crossing gives rise to a gapped spectrum which increasing linearly as we increase the angle with respect to the SL modulation direction. This induces the appearance of two Dirac cones. With a further increase in the SL strength, the two 
band crossing points move outside MBZ and the spectrum gets gapped out (see top figures of Fig. \ref{BLGSL}).

In an electric field SL, the situation is drastically different. Now, the SL potentials on the two layers are exactly opposite in sign, which can effectively be viewed as a periodic arrangement of potential kinks and anti-kinks. Earlier results have indicated that two topological 1D modes will be confined to each kink or anti-kink.\cite{Killi2,Martin,Xavier} These topological modes propagate perpendicular to the SL direction and have opposite velocities at a kink and an anti-kink. When one has such modes arranged periodically, they will couple each other and zero energy and finite energy Dirac cones will appear, as explained by an effective ``wire'' model in \Ref{Killi1} (see bottom figures of Fig. \ref{BLGSL}).

\subsection{Magnetic properties of bilayer graphene superlattices}

From the previous section, we can see that the phenomenon resulting from the existence of Dirac points is most manifest when magnetic field is weak compared to the SL strength. A very special feature of bilayer graphene SL is that anisotropic Dirac points can be generated in the energy spectrum, where originally only quadratic band touching points is present. Our numerical results also indicate that in intermediate magnetic field, the LLs are only slightly dispersive, compared to monolayer graphene. Therefore, in this section, we will focus on the properties of bilayer graphene SLs in weak magnetic fields using a two-band effective Hamiltonian. We hope that our results can provide means to the determination and characterization of these anisotropic Dirac points.
\begin{figure}
	[t] 
	\includegraphics[width=.48
	\textwidth]{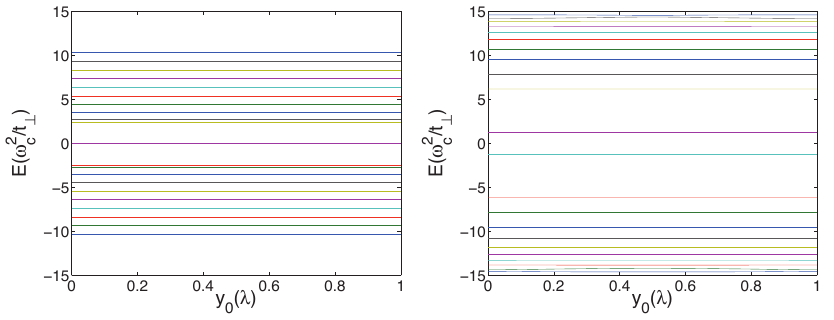}  \caption{(Color online) Energy spectrum of BLG subject to a chemical potential SL and a weak perpendicular magnetic field ($\ell_B=2\lambda$). For both panels, $\lambda=100a$, where $a=1.42\text{\AA}$. In the left panel, $V_0=0.01t$, two anisotropic Dirac points are generated in the spectrum in the absence of magnetic field, and the zero energy LLs are doubly degenerate. In the right panel, $V_0=0.04t$, no Dirac points are present, and no zero energy LL exists. In both panels, the energy bands are nearly flat. For both intermediate fields (not shown here), the energy bands are all slightly dispersive, in contrast to MLG SLs where energy bands overlap. This can be seen from the diagonal conductivities in Fig. \ref{BLGSL_diagonal} } \label{BLGSL_LL} 
\end{figure}

The low energy physics of bilayer graphene near the Brillouin zone corners at $\pm \bK$ is, in the presence of a perpendicular magnetic field, described by the following two-band effective Hamiltonian, \beq \hat{H}_0=-\frac{v_F^2}{t_{\perp}}\left( 
\begin{array}{cc}
	0 & (s \pi_x + i\pi_y)^2 \\
	(s \pi_x - i\pi_y)^2 & 0 
\end{array}
\right), \eeq with $s=\pm 1$ corresponding to $\pm \bK$ as in the case of monolayer graphene. Similar to the single layer case, we have replaced the momentum operator with its canonical counterpart to take into account of the magnetic field effect. In the following, we will work with the same gauge choice, $A=By\hat{x}$. For $s=+1$, the eigenvalues and the corresponding eigenvectors of the above Hamiltonian are \be &&\varepsilon_n={\rm sgn}(n)\sqrt{|n|(|n|-1)}\omega_c^2/t_{\perp},\nonumber\\
&&\phi_{n,k,+}(x,y)=\frac{e^{i k x}}{\sqrt{2 L}} \left( 
\begin{array}{c}
	\psi_{|n|,k}(y) \\
	- {\rm sgn}(n)\psi_{|n|-2,k}(y) 
\end{array}
\right), 
\label{Eq10}\ee with $|n| \geq 2$. In addition, there are two zero energy solutions, \be &&\varepsilon_1=0,\ \ \ \ \ \phi_{1,k,+} (x,y)=\frac{e^{ i k x}}{\sqrt{L}}\left( 
\begin{array}{c}
	\psi_{1,k} (y) \\
	0 
\end{array}
\right), \nonumber\\
&&\varepsilon_0=0,\ \ \ \ \ \phi_{0,k,+} (x,y)=\frac{e^{ ikx}}{\sqrt{L}}\left( 
\begin{array}{c}
	\psi_{0,k} (y) \\
	0 
\end{array}
\right). \ee For $s=-1$, the corresponding eigenvectors are given by $\phi_{n,k,-} (x,y) = \sigma_x \phi_{n,k,+} (x,y)$. The full low energy LL wavefunctions thus take the form $\phi_{n,k,\pm} {\rm e}^{\pm i K_x x}$. and these serve as a good basis to study the magnetic field effect of bilayer graphene SLs. 

The SL can be modelled by assuming spatially varying potential profiles on each layer, \beq \hat{H}_{sl}=\left( 
\begin{array}{cc}
	V_1({\bf x}) & 0 \\
	0 & V_2({\bf x}) 
\end{array}
\right), \eeq where different combination of $V_1$ and $V_2$ gives rise to different types of SL, e.g., $V_1({\bf x})=V_2({\bf x})$ for chemical potential SL and $V_1({\bf x})=-V_2({\bf x})$ for electric field SL. In following, we will only consider 1D SLs, where for proper choice of SL strength, various anisotropic Dirac points can be generated in the spectrum.

\subsubsection{Chemical potential superlattice}
\label{section:chemical}

\begin{figure}
	[t] 
	\includegraphics[width=.48
	\textwidth]{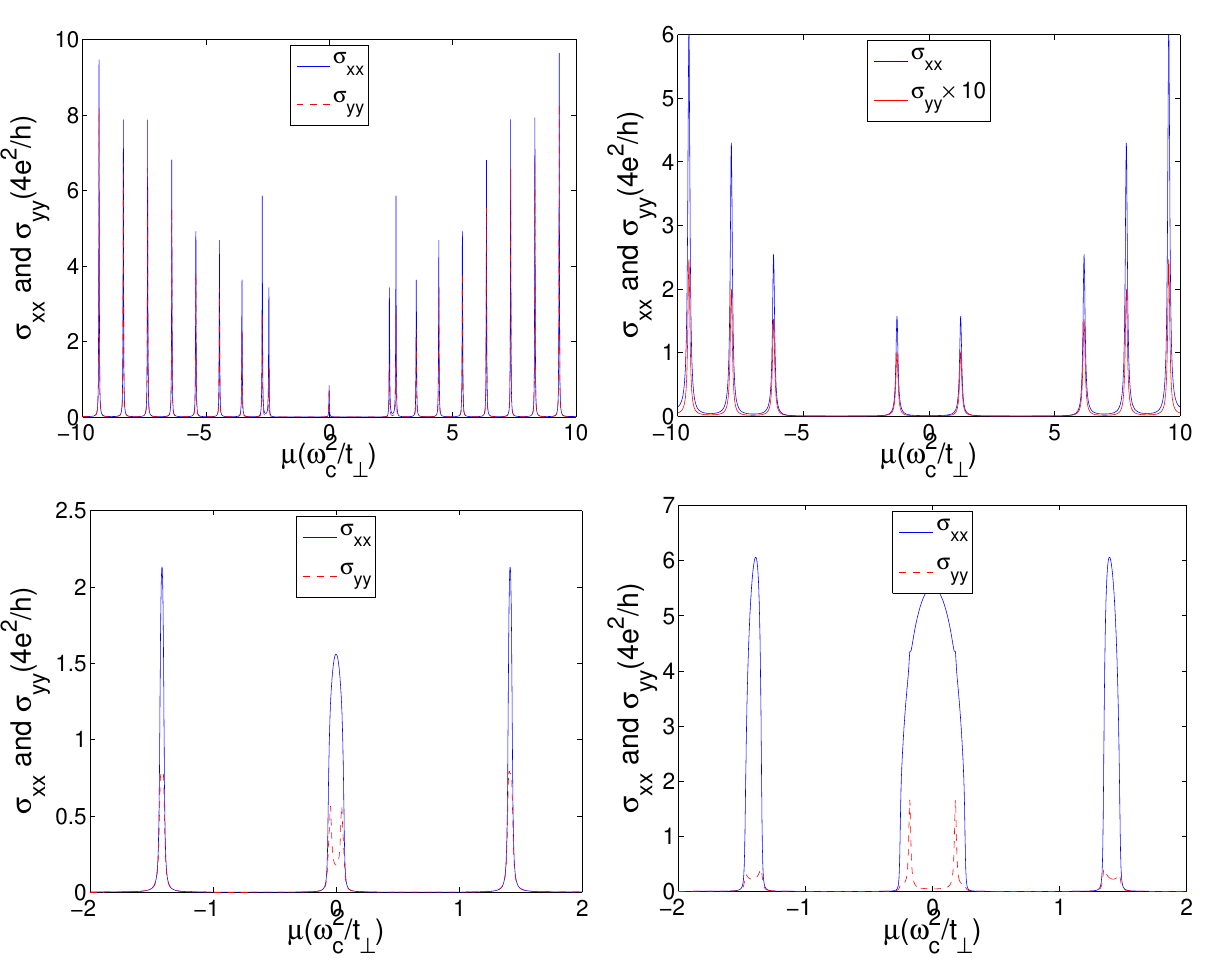}  \caption{(Color online) Diagonal dc conductivities of bilayer graphene SL for different SL strengths $V_0$, and magnetic fields $B$. The conductivity is shown for weak field ($\ell_B=2\lambda$, top panels) and intermediate field ($\ell_B=0.2\lambda$, bottom panels). Left panels (a,b) correspond to $V_0=0.01t$, and right panels (c,d) correspond to $V_0=0.04t$. The conductivities show anisotropy, where $\sigma_{xx}$ is always larger than $\sigma_{yy}$, in contrast to anisotropy reversal in MLG SLs. } \label{BLGSL_diagonal} 
\end{figure}

For 1D chemical potential SLs, we will take $V_1({\bf x})=V_2({\bf x})=V(y)=\frac{V_0}{2}\cos(2\pi y/\lambda)$. In previous work, \cite{Killi1} we have shown that for 1D chemical potential SL, two anisotropic Dirac points will derive from the original quadratic band touching points as long as the SL strength is smaller than a critical value. This results from the chiral symmetry protected level crossing in the SL direction and level anti-crossing in other directions. Once SL strength increases beyond this critical value, Bragg reflection will open up a gap in the energy spectrum on the mini Brillouin zone boundary. This behavior has also been found by Tan {\it et al}. \cite{Tan} 

Results for energy spectrum of chemical potential SL in a weak magnetic field are shown in Fig. \ref{BLGSL_LL}. In the left panel, we have chosen $V_0=0.01t$, where $t=3$~eV is the intralayer nearest neighbor transfer integral. For this value of SL potential, our previous calculation has shown that there will be two Dirac points generated in the mini Brillouin zone. Here, similar to the single layer case, the energy spectrum shows flat band structure. At the Dirac point, there are two nearly degenerate zero energy levels. Right above and below the Dirac point, we can also observe two sets of doubly degenerate energy levels. These are the $n=\pm 1$ LLs of each anisotropic Dirac cone. Moving further away from the Dirac point, no such degenerate levels are present. This is because these Dirac points only have linear dispersion at rather low energy. In the right panel, $V_0=0.04t$, energy gap opens up and there exist nonzero energy levels near the Dirac point. Energy bands in intermediate field regime (not shown here) are only slightly dispersive, where the amplitudes of dispersion are extremely small compared to the LL spacings. 

Fig. \ref{BLGSL_diagonal} shows the dc diagonal conductivity of chemical potential SLs. Similar to the single layer case, we can observe the anisotropy in the conductivity. However, there is no anisotropy reversal as magnetic field strength is tuned. In weak field, the transport anisotropy is determined by the anisotropic Dirac cones. From Ref. \onlinecite{Killi1}, $v_x \simeq 2v_y$ for emergent Dirac cones. Therefore, the conductivity in the $\hat{x}$ direction, $\sigma_{xx}$, should be larger than $\sigma_{yy}$ in a weak magnetic field. For intermediate magnetic field, due to the dispersion of the energy bands, the average velocity in the $\hat{x}$ direction is not zero, $\langle \hat{v}_x\rangle \neq 0$. On the other hand, $\langle \hat{v}_y\rangle$ is always equal to zero. This means that $\sigma_{xx}$ will acquire intra-LL contributions, while $\sigma_{yy}$ is mainly determined by inter-LL contributions and is thus small compared to $\sigma_{xx}$.
\begin{figure}
	[t] 
	\includegraphics[width=.48
	\textwidth]{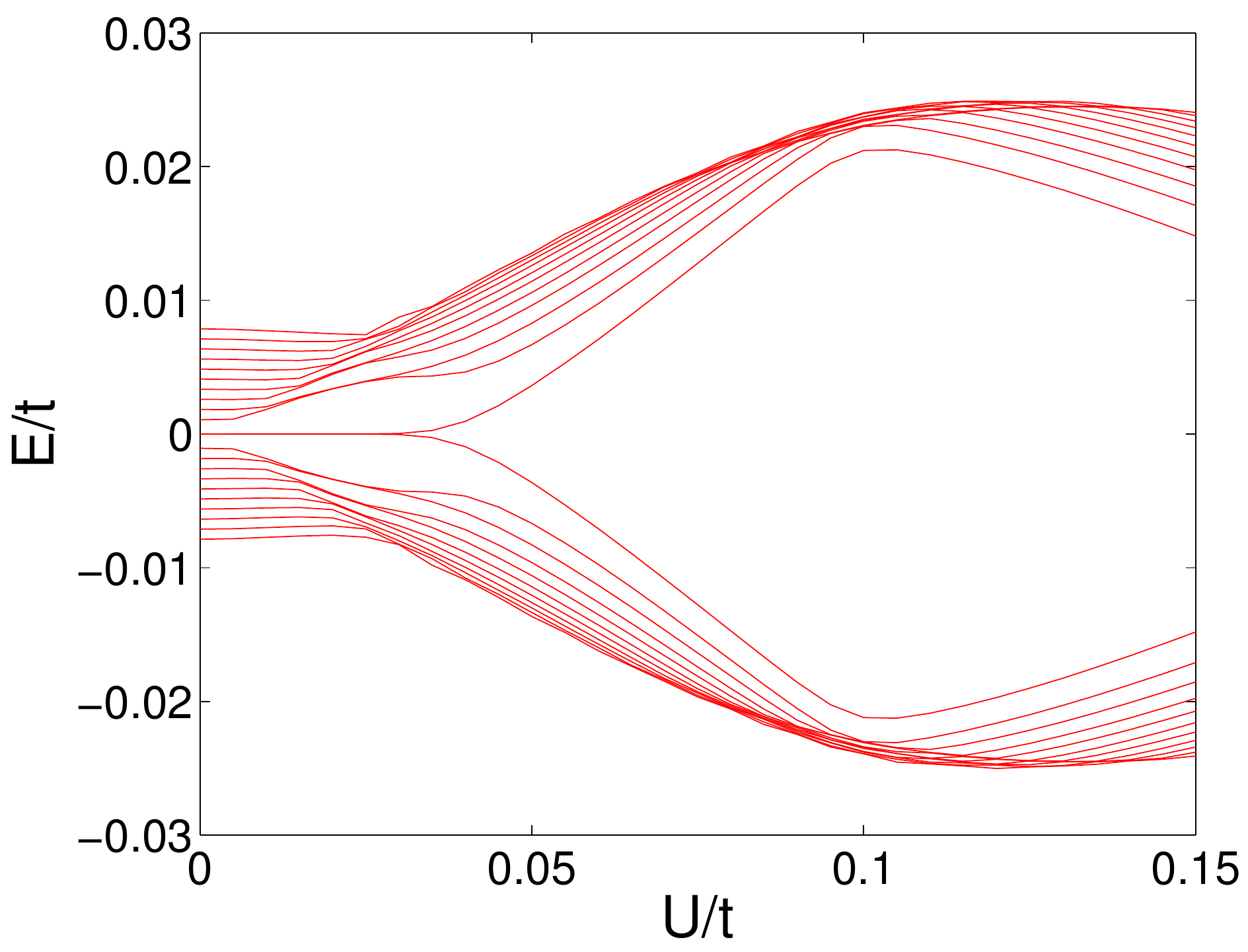} \caption{Evolution of low lying energy levels in a chemical potential SL as a function of SL potential strength $U$, with $\ell_B=2\lambda$, $y_0=0$, $\lambda=100a$ where $a=1.42\text{\AA}$. } \label{evolution} 
\end{figure}

We can further demonstrate how the flat energy levels evolve as the strength of the SL potential varies, as shown in Fig. \ref{evolution}. Here, we have fixed the magnetic field strength by setting $\ell_B=2\lambda$ and also fix the center position of the wavefunctions to be at 0. As the SL potential tends to zero, where the problem reduces to pristine bilayer graphene in a perpendicular magnetic field, we can see the energies of LLs will follow the pattern of Eq. (\ref{Eq10}). As SL potential increases, the physics gradually becomes dominated by anisotropic Dirac cones, which can be seen from the appearance of doubly degenerate levels at nonzero energies. This crossover from non-relativistic to relativistic behavior can be qualitatively understood by considering the competition between the characteristic energy scales in these two regimes. In pristine bilayer graphene, the low energy excitation are massive electron with an effective mass $m^*=t_{\perp}/2v_F^2$. In a magnetic field $B$, the characteristic energy scale is the cyclotron frequency, $\omega_c^{\prime}=eB/m^*c=1/m^*\ell_B^2$. On the other hand, the anisotropic Dirac points generated by SL have anisotropic Fermi velocities $v_y=\sqrt{2}\lambda|U({\bf Q})|/\pi$ and $v_x=2v_y$, where ${\bf Q}=\hat{y}2\pi/\lambda$. \cite{Killi1} Therefore, the characteristic energy scale associated with these Dirac points is $\omega_c=\sqrt{2v_xv_y}/\ell_B$. We expect to see a smooth crossover as these two energy scales are comparable to each other. A rough estimate shows that the crossover should occur around $U\sim 0.002t$, which is quite close to the value that we have observed.

Further increasing the SL potential, the doubly degenerate zero energy levels become gapped and all levels are pushed away from Dirac point. Surprisingly, at rather strong SL potential, $U\sim 0.22t$, zero energy LLs appear again, and all of the higher energy levels become doubly degenerate. This phenomenon can be understood from the result of Tan {\it et al}. \cite{Tan} As it has been shown, for a chemical potential SL, when SL potential is strong enough, anisotropic Dirac cones will show up again in the energy spectrum, which naturally leads to the zero energy LL in the presence of a magnetic field. According to Ref. \onlinecite{Tan}, for certain values of SL potential, there can be four Dirac points in the spectrum, which is not present in our calculation. Of course, for strong SL potential, the two-band description of bilayer graphene is no longer valid. However, we have reproduced this result in a four-band effective Hamiltonian approach. In strong chemical potential SL, there can exist four-fold degenerate zero-energy LLs, and the degeneracy 
reduces to two upon further increasing the SL strength.

\subsubsection{Electric field superlattice}\label{section:electric}

For 1D electric field SLs, we will take $V_1({\bf x})=-V_2({\bf x})=V(y)=\frac{V_0}{2}\cos(2\pi y/\lambda)$. Following the same procedure as in the previous section, we have calculated the energy spectrum and the dc conductivities for an electric field SL in a magnetic field. 
\begin{figure}
	[t] 
	\includegraphics[width=.48
	\textwidth]{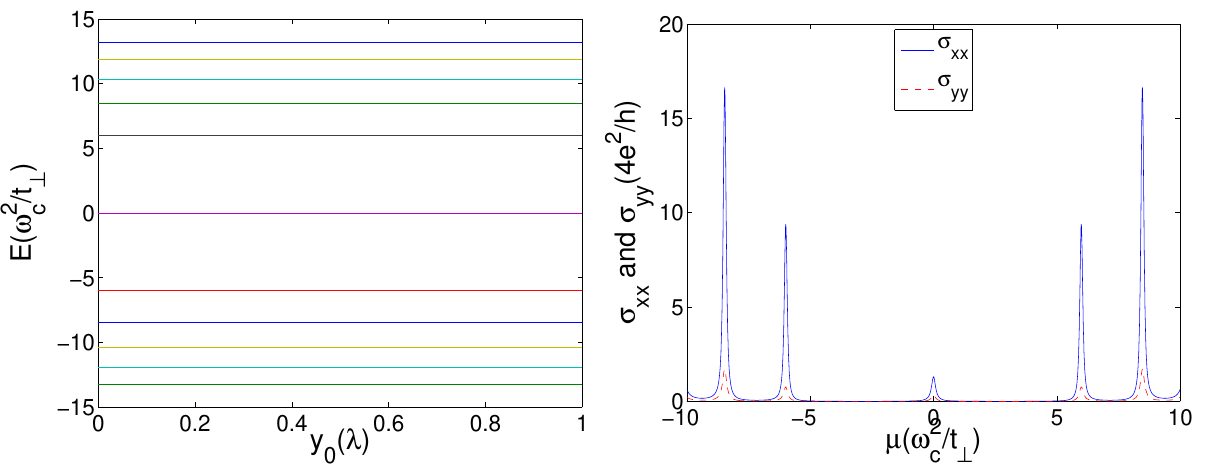}  \caption{Energy spectrum and dc diagonal conductivities of electric field superlattice, in a weak magnetic field, with $\ell_B=2\lambda$, $U=0.04t$, $\lambda=100a$ where $a=1.42\text{\AA}$. Results for intermediate magnetic fields are similar to those of chemical potential SLs and are not shown here.} \label{delta} 
\end{figure}

\begin{figure}[t] 
	\includegraphics[width=.48
	\textwidth]{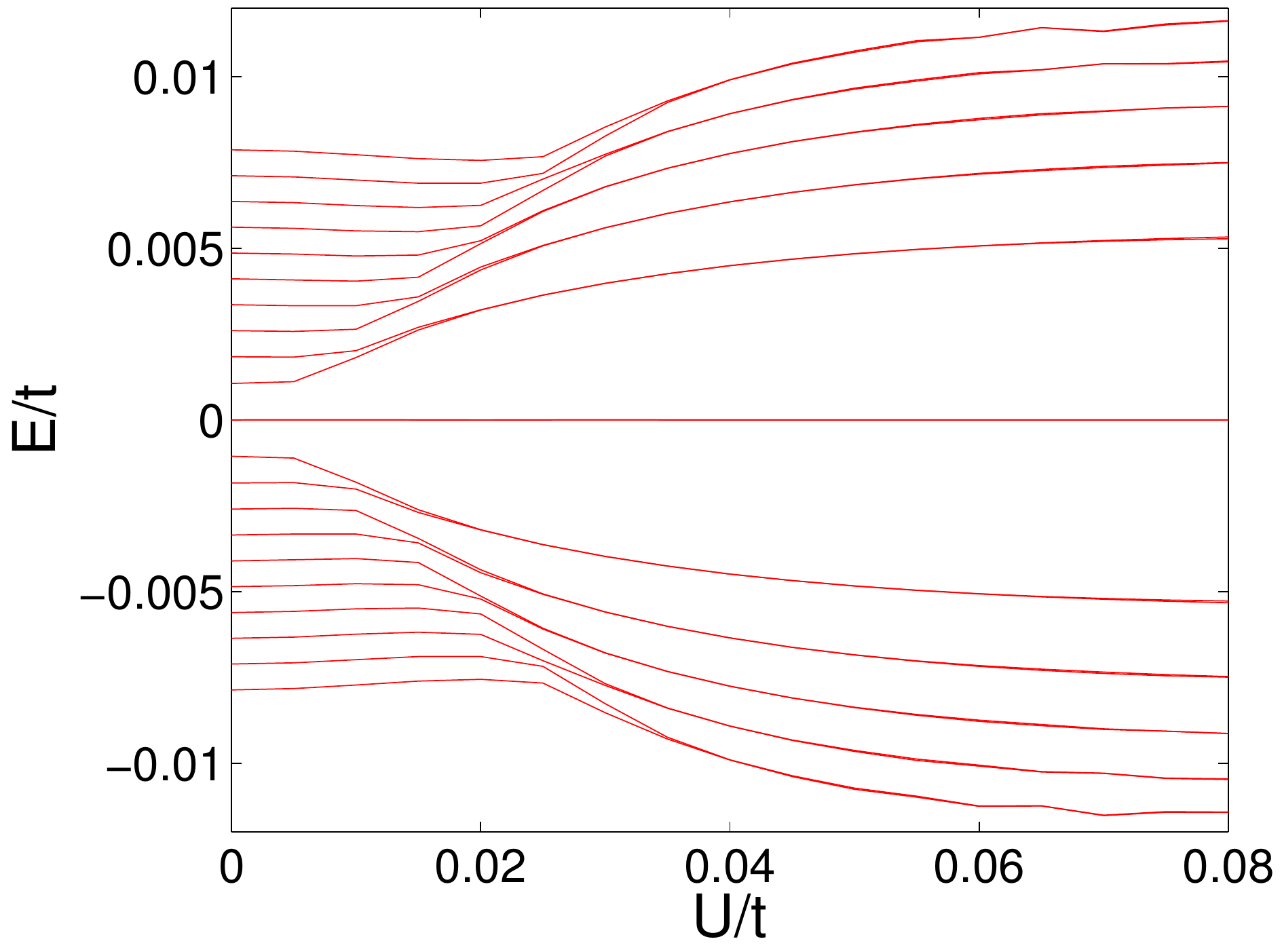} \caption{Evolution of low lying energy levels in an electric field SL as a function of SL potential strength $U$, with $\ell_B=2\lambda$, $y_0=0$, $\lambda=100a$ where $a=1.42\text{\AA}$.} \label{evolution_eSL} 
\end{figure}

In a 1D symmetric electric field SL, there are always two zero energy Dirac points present in the spectrum, which results from the coupling of 1D zero modes at kink/antikink of the SL potential profile. Earlier, we have developed an effective ``wire'' model to describe these Dirac points and demonstrated that they are topological stable if a generalized ``inversion'' symmetry, $V(y+\lambda/2)=-V(y)$ ({\it i.e.}, flipping the electric field and shifting $y$ by $\lambda/2$), is preserved.\cite{Killi1} This implies when a weak magnetic field is applied, doubly degenerate zero energy levels should appear at the Dirac point. Indeed, as it can be seen from Fig. \ref{delta} and \ref{evolution_eSL}, these two levels always stay at zero energy, independent of the SL potential. Moreover, for the SL potential we choose, $V_0=0.03t$, nearly degenerate levels appear even at nonzero energies (Fig. \ref{delta}). These correspond to the LLs derived from anisotropic Dirac points, up to $n=\pm 4$. From Fig. \ref{evolution_eSL}, it is more clear that at strong SL potential, physics is strongly dominated by the Dirac points, where higher energy levels become doubly degenerate and resemble the higher LLs of the Dirac cones. When SL potential is weak, equally spaced LLs are recovered, as in the chemical potential SL, which also indicates a non relativistic to relativistic crossover at certain SL strength. Remarkably, and different from the chemical potential SL case, the relativistic behavior survives to higher energies as SL potential increases, which means the linear approximation description of Dirac cones works in a larger energy range. This is consistent with earlier result. \cite{Killi2} Consider a potential profile with a kink of bias reversal. When the bias is large, the conduction and valence bands are widely separated and a pair of linearly dispersed zero modes will traverse the gap. Now, when these kinks and antikinks are arranged periodically, Dirac cones will result from the coupling of these modes, and the Fermi velocity along SL direction inherits its value from the freestanding zero mode. \cite{Killi1} Therefore, as the SL potential increases, the energy range where the Dirac cone approximation is valid also increases, which leads to the robust relativistic physics at large SL potential.

\section{Real space picture for interlayer bias modulations in bilayer graphene}
\label{section:realspace}

As suggested in an early study by the present authors \cite{Killi1}, it is extremely useful to view an interlayer bias modulation as establishing a series of `kink' and `antikink' modes along the zero-lines where the interlayer bias reverses polarity. Within the effective $2\times 2$ low-energy description of BLG, the interlayer bias can be regarded as a mass generating term and the zero-lines represent boundaries where the sign of the mass changes. As consequence, along a single isolated kink (antikink), a pair of two right (left) moving `topological' modes at $\bK$ and corresponding two left (right) moving modes at the other $-\bK$ \cite{Martin, jung, qiao, Killi1}. The typical length scale over which these kink-antikink modes are confined, $l$, varies from $\sim100d$ for $\Delta V=0.01t$ to $<50$ lattice sites for $\Delta V=0.1t$, where $d=\sqrt{3}a$ is the length the primitive lattice vector.  For the remainder of this section we set $d=1$.

Accordingly, an interlayer bias modulation can be thought of as a periodic array of coupled kink-antikink solitons. With this viewpoint, and using the symmetry properties of the kink and antikink modes, Killi {\it et al.}\ \cite{Killi1} derived an effective low-energy hamiltonian describing how anisotropic Dirac cones precipitate precisely at the band crossing point between the kink and antikink modes when the soliton modes overlap and couple. With this same perspective, further intuition into the quantum Hall effect in the SL system can be obtained by studying the properties of 1D kink states in a magnetic field.

Aside from its relevance to SLs, a study into the properties of the kink/antikink states also provides valuable insights into disorder BLG samples in the quantum Hall regime. Various sources of disorder are known to generate a random electrostatic potential landscape. Throughout these samples, 1D kink-states are expected to percolate along precipitous fluctuations that reverse the parity of interlayer bias. If the percolation networks are well extended, these kink states would contribute to the breakdown of the quantum Hall effect \cite{connolly}. Recent STS measurements of BLG samples in the quantum Hall regime also indicate that even when the sample is uniformly biased by external gates, the disorder is still strong enough to locally reverse the polarity of the interlayer bias \cite{rutter, abergel}. These recent measurements provide an additional motivation to study the properties of kink states in a magnetic field and, specifically, to examine how they modify the tunneling current.

In this section, we consider interlayer bias modulations in quantum Hall regime from the perspective of the kink-antikink modes. We study the dilute limit where the kink and anti-kink states are well separated and confined to the zero-lines (i.e.~$\lambda \gg l$). Within this regime, the low energy states of the system are completely dominated by midgap soliton modes. We start by first reviewing the effects of a magnetic field on uniformly biased BLG in section \ref{section:uniform} and then non-uniformly biased BLG with a single isolated anti-kink in the potential profile in section \ref{section:kink} . In addition to studying the dispersion relation, we examine how the tunneling current is modified in the proximately of a kink. Next, we consider an array of decoupled kink and anti-kink modes in Section \ref{section:array} . For magnetic fields such that $\ell_B < \lambda$, it is only necessary to understand the dispersion relation of an isolated kink/anti-kink pair as this is sufficient to describe the entire band structure. This allows us to consider a zigzag ribbon of width $N_{cell}$ with one period of modulation (i.e.~a pair of kink anti-kink states) and implement a Peierl's substitution in the full tight-binding Hamiltonian. For weaker magnetic fields such that $\ell_B \gtrsim \lambda$, it becomes necessary to include a larger number of kink/anti-kink states to describe the high energy features of this system. Despite this, further insights into the low energy modes can be made by combining the knowledge obtained from the study a single kink/anti-kink pair and by using a simple low energy effective theory (see Ref. \onlinecite{Zarenia1}). Aside from this, in section \ref{section:valley} we demonstrate that a coupled pair of kink/anti-kink states opens asymmetric bandgaps in each valley that can be precisely controlled using a magnetic field. This effect may provide a viable route to developing a switchable valley filter.

\subsection{Uniform Bias} 
\label{section:uniform} 

Although it has been discussed extensively in the literature, we review and emphasize a few important points about the LL structure in uniformly biased BLG. Details about the LL structure in biased BLG can be found in Ref.~\onlinecite{McCann1, McCann2, castro2007, pereira, nakamura, mazo2011, nakamura2, nakamura3}; here we only briefly summarize some of the features as they pertain to our current discussion.
\begin{figure}
	[t] 
	\includegraphics[width=8.3cm]{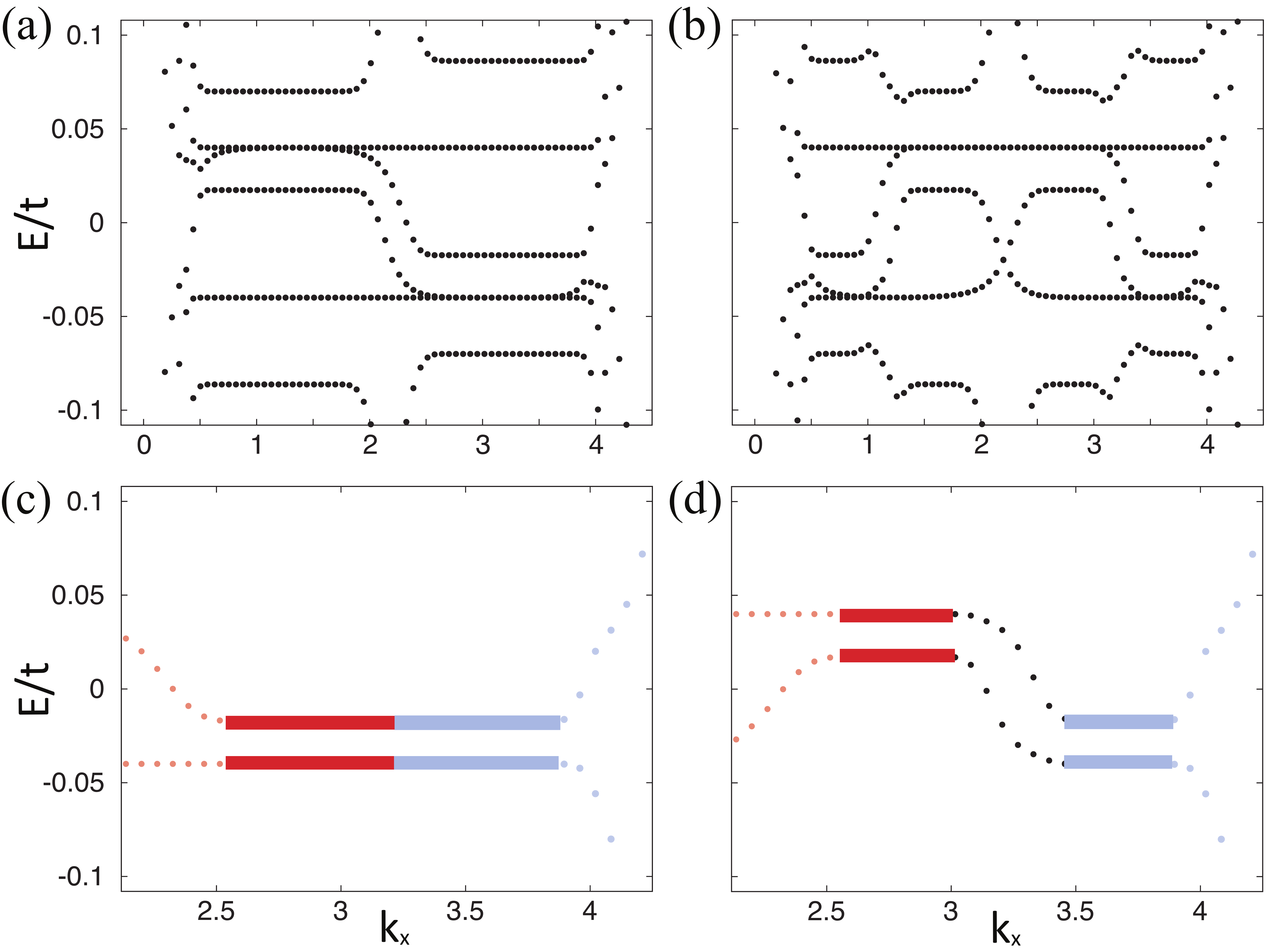}
	\caption{Landau Level spectrum of graphene strip in the presence of (left column) a uniformly interlayer bias and (right column) with an anti-kink bias profile with $\Delta V=0.08t$, $N_{cell}=300$. (c,d) n=0 and n=1 LLs. Solid red (blue) lines are bulk states localized in the upper (lower) half of the sample and red (blue) dotted lines are the upper (lower) edge states. Black dotted lines connecting the bulk LLs are anti-kink states. We have chosen a strong magnetic field of $B=120 T$ in order to make the relevant features easily discernible and note that the spectrum is qualitatively identical at lower magnetic fields.} \label{KinkEdge} 
\end{figure}

In the Landau gauge that preserves translation symmetry along the zigzag direction, inspection of Fig.~\ref{KinkEdge} (a) shows that there are two distinct energy level spectrums in the dispersion relation, one about each valley. Each energy level consists of two well-defined regions: a flat, non-dispersive region consisting of bulk LLs and dispersive edge states \cite{mazo2011, nakamura2} (note, the edge state dispersion for the $n=0$ and $n=1$ are unique in that they contained non-dispersive regions, as seen in Fig.~\ref{KinkEdge} (a,c)). The wavefunctions of the non-dispersive LL states are well confined within the bulk and, starting from states on the far left side of the level and ending on the right side, move from one edge of the sample to the other.

A comparison of the spectrum about the \bK-points clearly shows that the interlayer bias breaks the valley degeneracy of the LL spectrums \cite{McCann2, nakamura}. Although the valley degeneracy is broken, the states in each valley are connected by a symmetry that relates the low energy Hamiltonian of opposite $\bK$-point. If we define the operator $\cal{P}$ to correspond to the interchanging of $A-B$ sublattices combined with layer exchange (i.e. $A_1\leftrightarrow B_2$ and $B_1\leftrightarrow A_2$), it is a simple matter to show that ${\cal P}^\dag H_\bK(-\Delta V){\cal P}=H_{\bK'}(-\Delta V)$. Thus, upon interchanging the sites labelling, the quasiparticles in each valley are governed by the same low energy Hamiltonian but with the effective relative parity of the interlayer bias reversed, leading to the observed valley degeneracy breaking.

Inspection of the band structure shown in Fig.~\ref{KinkEdge} (c) reveals that in the presence of an interlayer bias, the zero energy $n=0$ and $n=1$ LLs are shifted to finite energy and their degeneracy is lifted. Note, the energy shift of these two these states is positive in one valley an negative in the other. A second crucial observation about the $n=0$ and $n=1$ Landau states (and to a lesser extent the higher energy states) is that the wavefunctions are strongly localized to one of the two layers determined by its valley index. Such layer polarization is directly observed in STS measurements and information about the LL spectrum, valley index and local interlayer bias can be obtained \cite{rutter}. Below, we discuss further how the presence of a kink and anti-kink affects the tunneling current and give rise to clear signatures observable in STS measurements.

\subsection{Single Kink or Antikink in the Bias} 
\label{section:kink}

\begin{figure}
	[t] 
	\includegraphics[width=7.5cm,angle=0]{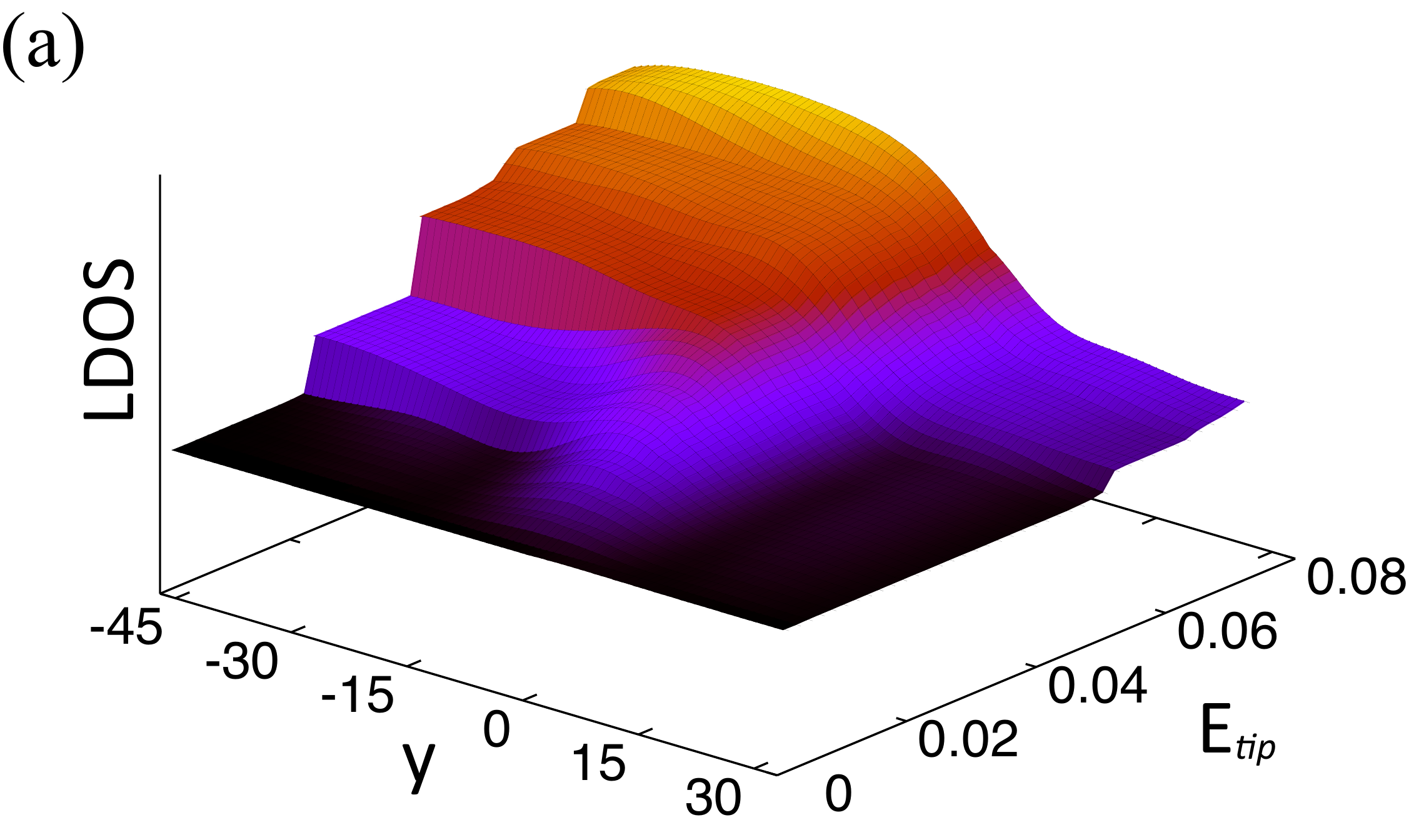} \\
	\vspace{0.1cm} 
	\includegraphics[width=8.5cm,angle=0]{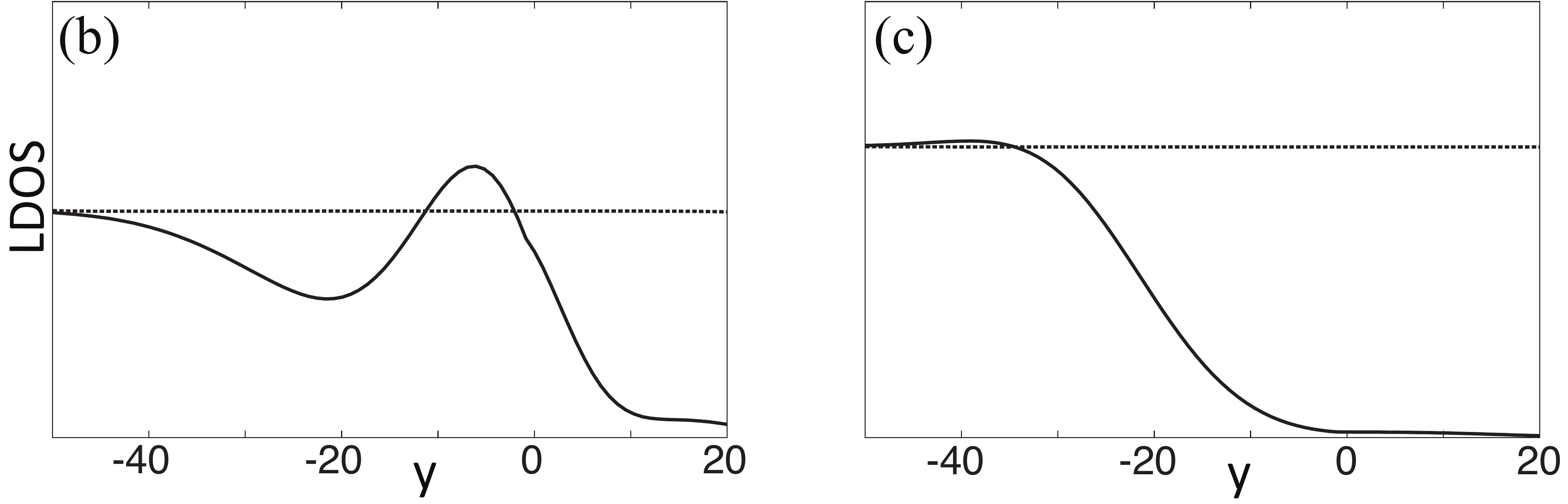} 
	
	\caption{Tunneling current along the $y$ direction (in units of $d$) in the proximity to kink in the interlayer potential bias with $|\Delta V=0.08t|$ and large $B=80 T$ to avoid extraneous finite size of affects. (a) Tunneling current across a kink for states as function of the sample-to-tip bias, 
$E_{tip}$, and $\mu=0.0$. (b) Cut where $E_{tip}$ is between the $n=0$ and $n=1$ LL ($E_{tip}=0.0.025t$). Shift of weight towards the interface is a signature of the kink modes. (c) Tunneling current with $E_{tip}$ above $n=0$ with $\mu$ in between the $n=1$ and $n=0$ LL 
($E_{tip}=0.042t$ and $\mu=0.037t$). Evidence of a kink state is given by the sharp suppression 
of the tunneling current, as Pauli blocking prevents tunneling into the kink states.} \label{LDOS} 
\end{figure}

In the absence of a magnetic field, an isolated kink in the interlayer bias generates two unidirectional dispersing subgap bands in one valley and two oppositely dispersing bands in the other valley related by time reversal. An anti-kink generates similar low energy bands, although the velocity of these modes is reversed in each valley (see Fig.\ \ref{KinkEdge} and Ref. \onlinecite{Martin} for details). 

When a large magnetic field is introduced, the energy spectrum resembles that of the LL structure of uniformly bias BLG, but with further distinctive features. Consider the energy spectrum about a single valley. Comparison of Fig.~\ref{KinkEdge} (c) and (d) shows that instead of each energy level having a single non-dispersive bulk LL, each LL breaks into two flat regions shifted in energy, which are connected through a dispersive mode. States in these two flat regions correspond trivially to the bulk Landau states that would have been generated by having a uniform interlayer bias of the same parity as the local bias. Hence, these two regions are composed entirely of Landau states that are well localized in the bulk and respond only the local interlayer bias. As in the previous case, far way from the ${\bK}$-point the bulk states evolve into edge states, and are again sensitive only to the local interlayer bias at that edge.

However, it is for the $n=0$ and $n=1$ LLs that the presence of the kink states are directly observed. Starting from a bulk state in the $n=0$ ($n=1$) Landau level on one side of the interface and tracking the states to the other side of interface (moving left to right in momentum space), we observe the following sequence. The bulk $n=0$ ($n=1$) LL states begin to continuously evolve into dispersive kink states as the guiding center approaches the zero-line and then emerge from the interface on the other side as bulk states in the $n=1$ ($n=0$) LL.

Although other dispersive states connecting the high energy bulk LL on either side of the interface are also present, they are not of topological origin. These states are simply LL states that becomes dispersive as their wavefunction begins to overlap with the region of spatially varying potential. Hence, the properties of these states are inherently sensitive to the the magnetic field and the details of potential profile. In contrast, the kink-states are of topological origin and so persist even in the absence of a magnetic field. Due the strong confined of these states, the properties of these states are remarkably robust and insensitive to the magnetic field (see Zarenia {\it et al.}\ \cite{zarenia} for more details).

\begin{figure} [tb] 
	\includegraphics[width=8.5cm]{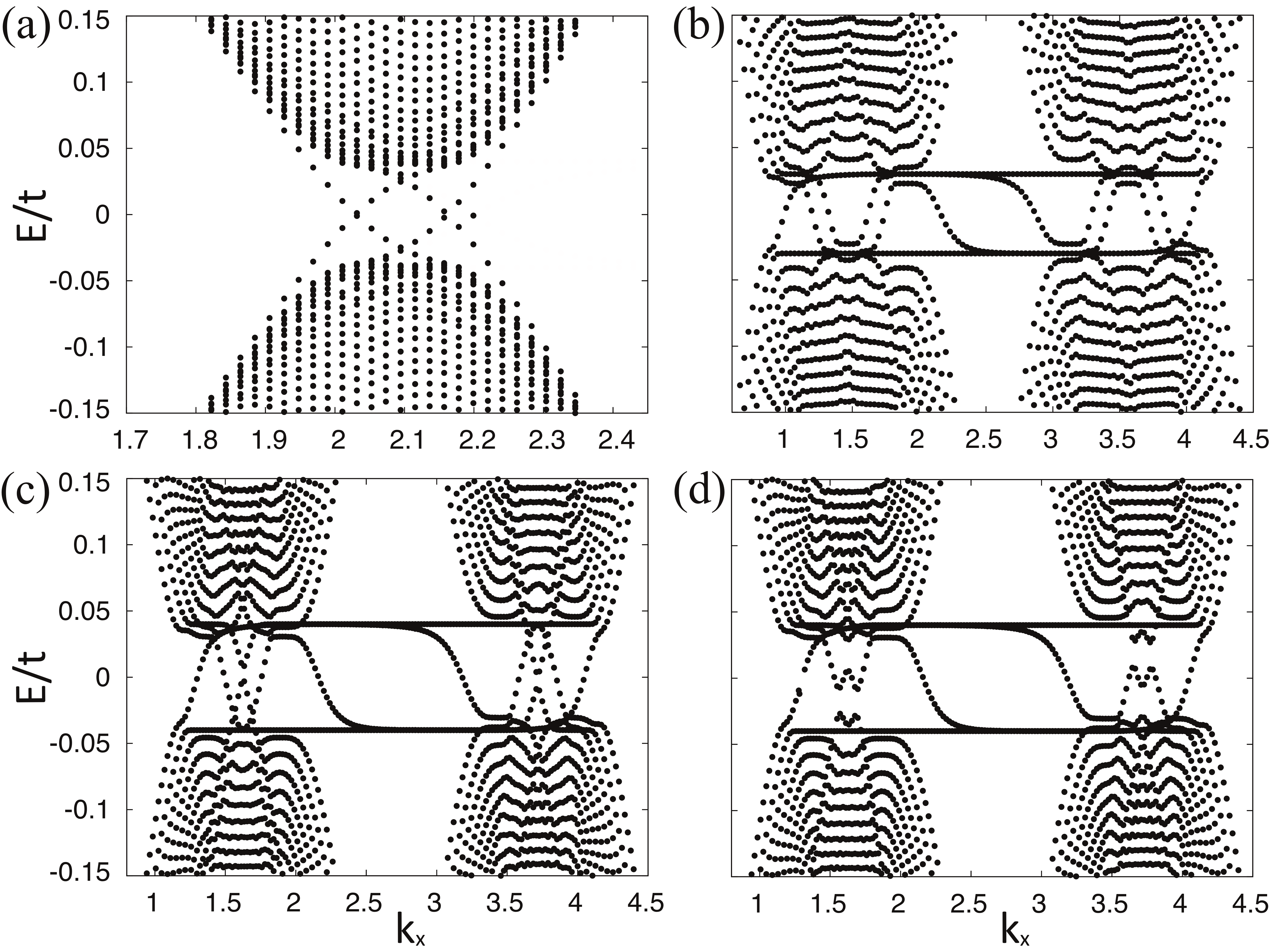} 
	\caption{Dispersion of BLG in the presence of kink/anti-kink with $|\Delta V|=0.08t$. (a) decoupled modes with $B=0T$, (b) decoupled modes with $\lambda_{sep}=100d$ and $\ell_B=17d$, (c) $\lambda_{sep}=40d$ and $\ell_B=17d$, and (d) $\lambda_{sep}=20d$ and $\ell_B=17d$.  Note, the large spread of the wavefunctions leads to coupling of the zero energy modes for $\lambda_{sep} < 40d$ and the qualitative features remain robust for similar $\ell_B/\lambda_{sep}$.}
	\label{Pair}
\end{figure}

The tunneling current of the $A_2$ sites in the vicinity of a kink is shown in Fig.\ \ref{LDOS}. It was computed by allowing electrons to tunnel within an energy window between the chemical
potential $\mu=0$ and the sample-to-tip bias $E_{tip}$. Away from the interface, sharp LL plateaus easily distinguish the LL spectrum. The marked suppression of the tunneling current on the other side of the interface where the parity of the bias is reversed is indicative of the strong layer polarizability discussed previously. Only when the $E_{tip}$ lies above the $n=2$ LL is there any weight on the $A_2$ sublattice in this region, consistent with the low energy theory.

When $E_{tip}$ lies between the chemical potential and the $n=1$ LL, there is notable enhancement of the tunneling current close to the interface. A cut along $E_{tip}=0.025t$ shown in Fig.\ \ref{LDOS} (b) shows signatures that this enhancement is due to the presence of strongly confined kink states. Namely, the confinement modes to the zero-line manifests as a transfer of weight from the bulk to the interface and results in the combined downturn in the bulk and sudden enhancement in the tunneling current close to the interface. Evidence of the strong localized states can also be seen when $E_{tip}$ lies just above the $n=0$ LL and $\mu=$ lies just below. Here, the tunneling into the kink states is suppressed by Pauli blocking, leading to a sharp reduction in the tunneling current close to the interface. In contrast, at higher energies the reduction in the tunneling current is a very gradual rolloff across the interface, showing no indication of localized interface modes.

Evidence demonstrating the robustness of these modes in the presence of magnetic field suggest these modes may play an important role in establishing percolation networks responsible for the break down of the quantum Hall effect.\cite{connolly}

\subsection{Array of Kinks and Antikinks} 
\label{section:array}

We begin by first considering the dilute limit where the kink and anti-kink soliton modes are well separated and decouple. In the the regime where $\ell_B < \lambda$, the LL states are well localized and respond only to their local environment. Hence, the dispersion relation is expected to consist of a series of bulk regions -- identical to those generated by a uniform bias with alternating bias parity -- connected by intermediate kink/anti-kink states. This viewpoint is confirmed by inspecting the dispersion relation of a single kink/anti-kink pair (which, neglecting the spurious edge states, compose the unit cell) shown in Fig.\ \ref{Pair} (b).


In the opposite regime where $\ell_B > \lambda$, the dispersion relation in Fig.\ \ref{Pair} (c) indicates the bulk LLs become dispersive in general. Here, the wavefunctions are spread over a wide region encompassing the kink/anti-kink of the potential profile, and thereby altering their properties. In contrast, with respect to the low energy kink/antikink states the magnetic field again does little more than shift their position in momentum space; the mode velocity is robust and the wavefunctions are rigid. Only at higher energies, close to the LL energy levels are the dispersive kink states modified.

In sum, the effect of a increasing magnetic field at low energy is to shift the kink and anti-kink modes relative to each other (the anti-kink modes move right the kink modes move left in momentum space). Since the mode velocity of the both the kink and anti-kink are reversed in each valley, this shift pushing the band crossing points up in energy in one valley and down in the other. In this way, a magnetic field can be used to control the band crossing points between modes of adjacent wires.

Further insights into the robustness of the low energy modes present in the continuum model of the 1D interlayer bias SL studied above can also be made by employing a simple low energy theory inspired by the kink/anti-kink viewpoint. In the dilute limit and in the absence of a magnetic field each kink (anti-kink) generates an identical copy of the low energy modes about each valley. At each \bK-point in Fig.~\ref{Pair}~(a), the two modes for each kink (antikink) can be linearized about zero energy (see Ref.~\onlinecite{Killi2} for more details). An effective low energy hamiltonian of the SL about one of the \bK-points can be written as 
\begin{eqnarray}
	H(\px)=v_0\sum_n\big{[}(-1)^n (p_x-p^*_1) c^{\dg}_{\px n}c_{\px n} \notag \\
	+(-1)^n (p_x-p^*_2) f^{\dg}_{\px n}f_{\px n}\big{]} 
\end{eqnarray}
where $v_0$ is the velocity of the modes about zero momentum, and $c^{\dg}_{\px n}$ ($f^{\dg}_{\px n}$) are operators that create electrons on wire $n$, with momentum $p_x$ about momentum close to one of the two zero energy crossing points labelled $p^*_{i}$. (Note, the low energy theory of the opposite valley is identical except that the velocity of the kink/anti-kink modes are reversed.) Representing a magnetic field by the Landau gauge field ${\bf A}=-By\hat{x}$, electrons hopping along a given wire $n$ see an average constant gauge field given by ${\bf A_n}=-B\lambda n$. In a magnetic field, the low energy effective Hamiltonian becomes 
\begin{eqnarray}
	H(\px)=v_0\sum_n\big{[}(-1)^n (p_x+n\lambda/\ell_B^2-p^*_1) c^{\dg}_{\px n}c_{\px n} \notag \\
	+(-1)^n (p_x+n\lambda/\ell_B^2-p^*_2) f^{\dg}_{\px n}f_{\px n}\big{]}. 
\end{eqnarray}
These results demonstrate that the full dispersion of the SL consists of multiple copies a single kink-antikink pair dispersion repeated every $\lambda/\ell_B^2$ in momentum space, consistent with the numerically computation of the dispersions relations provided above. Notice, that the modes in a given wire crosses modes in the two neighbouring wires symmetrically about zero energy. If the magnetic field is weak, the kink modes of wire $n$ cross with the anti-kink modes of wire $n\pm1$ at an energy of $\pm v_0 \frac{\lambda}{2\ell_B^2}$. If the modes couple, level repulsion will cause the kink/antikink modes of the neighbouring wire to mix and split, but because the bandcrossing is symmetric about zero energy and each the each mode couples to both neighbouring modes equally (for a symmetric SL), a zero energy band remains intact.

\subsection{Valley Filter} 
\label{section:valley} 

In studying a kink/anti-kink single pair in the coupled regime, we observe a remarkable effect not observed in a previous study \cite{zarenia} of the kink states that used a low energy model that did not capture the valley degeneracy lifting.  As the wavefunctions of the kink and antikink modes begin to overlap, a bandgap opens at the band crossing points.  Since the size of the bandgap is determined by the degree of overlap between the wavefunctions, it can be narrowed by either increasing the kinks-antikink separation or by reducing the interlayer bias, which causes the wavefunctions to spread further from the interface.  Furthermore, as the magnetic field strength increases, the bandgap shifts to positive energy in one valley and negative energy in the other due to the valley asymmetry in the band crossing point. Remarkably, for a single coupled kink/anti-kink pair an energy window opens where only unidirectional modes are present along each kink and anti-kink (the direction of flow can be flipped by reversing the magnetic field). Moreover, all the conducting modes have the same valley index and so acts as a valley filter similar to those suggested along edge states\cite{Rycerz} and line defects \cite{gunlycke, wright} when the chemical potential is tuned to lie in the bandgap.

\section{Summary and discussion} 

In this paper, we studied magnetic properties of MLG and BLG under various types of 1D external potentials.  For MLG, we identified three regimes of magnetic field strengths that generate distinctive features in the LL dispersion and transport properties.  Under a weak magnetic field, MLG SLs exhibit zero energy LLs whose degeneracies are identical to the numbers of Dirac points present in the spectrum.  At higher energies but still within the linear range of the spectrum, differences between the Dirac cones cause the LL degeneracy to be lifted.  Therefore, measurements that carefully determine the degeneracy of the LLs can be used to probe and characterize the underlying anisotropic Dirac cones. We further showed that the diagonal conductivities show strong anisotropy, with conductivity in the SL direction larger than that in the transverse direction. Interestingly, the anisotropy can be reversed for an intermediate magnetic field, where the LLs become dispersive. This field tunable anisotropy may find interesting device applications such as in switching or in resistive bits.

We then considered two types of SLs for BLG and again showed that analysis of the LL spectrum provides clear signatures of the underlying Dirac cones. However, the effects of a magnetic field were shown to be rather different in the bilayer SL systems than on the monolayer SL system.  Although the diagonal conductivity of both SLs also have strong anisotropy like in the monolayer, there is no field tunable anisotropy reversal.  Furthermore, regardless of the type of SL there are distinct crossovers from nonrelativistic physics to relativistic Dirac physics as the SL strength is tuned.  In the case of a chemical potential modulation, the zero energy LL vanishes at a critical SL strength where the Dirac cones becomes gapped. As for the electric field SL, two zero energy LLs are always present irrespective of the strength of the magnetic field SL strength or magnetic field.  This demonstrates a remarkable robustness of the Dirac points inherited from their topological origin.

We have also studied how a magnetic field affects BLG subject to a uniform bias and with bias reversing kinks in the potential profile.  The motivation here was to gain a real space perspective of the LL wavefunctions in the quantum hall regime.  Together with a low energy model, this study further established an intuition into the properties of the electric field SL.  Moreover, it provided valuable insights into the topological kink states present in disorder BLG and suggests how that they may contribute to the breakdown of the quantum Hall effect.  Finally, we proposed that a pair of coupled kink-antikink modes subject to a magnetic field could serve as a possible route to fabricate a switchable one-way valley filter.

\begin{acknowledgements}

We acknowledge Gil Refael, Jeil Jung, and Jin-Luo Cheng for useful disscussions.  Financial support was provided by the 
Early Research Award, Government of Ontario, NSERC of Canada and a University of Waterloo start-up grant.

\end{acknowledgements}

\end{document}